\def \beq {\begin{equation}}
\def \eeq {\end{equation}}
\def \beqa {\begin{eqnarray}}
\def \eeqa {\end{eqnarray}}
\newcommand{\mc}[1]{\multicolumn{#1}}   
\newcommand{\dr}[1]{_{\rm #1}}
\newcommand{\ur}[1]{^{\rm #1}}
\newcommand{\req}[1]{(\ref{#1})}
\newcommand{\vect}[1]              
           {\mbox{\boldmath$#1$}}  
\documentclass[12pt]{article}

\makeatletter
 \@addtoreset{equation}{section}
 \def\theequation{\thesection.\arabic{equation}}
\makeatother
\makeatletter
\newcount\@tempcntc
\def\@citex[#1]#2{\if@filesw\immediate\write%
                  \@auxout{\string\citation{#2}}\fi
  \@tempcnta\z@\@tempcntb\m@ne\def\@citea{}\@cite{\@for\@citeb:=#2\do
    {\@ifundefined
       {b@\@citeb}{\@citeo\@tempcntb\m@ne\@citea%
                   \def\@citea{,}{\bf ?}\@warning
       {Citation `\@citeb' on page \thepage \space undefined}}%
    {\setbox\z@\hbox{\global\@tempcntc0\csname b@\@citeb%
                     \endcsname\relax}%
     \ifnum\@tempcntc=\z@ \@citeo\@tempcntb\m@ne
       \@citea\def\@citea{,}\hbox{\csname b@\@citeb\endcsname}%
     \else
      \advance\@tempcntb\@ne
      \ifnum\@tempcntb=\@tempcntc
      \else\advance\@tempcntb\m@ne\@citeo
      \@tempcnta\@tempcntc\@tempcntb\@tempcntc\fi\fi}}\@citeo}{#1}}
\def\@citeo{\ifnum\@tempcnta>\@tempcntb\else\@citea\def\@citea{,}%
  \ifnum\@tempcnta=\@tempcntb\the\@tempcnta\else
   {\advance\@tempcnta\@ne\ifnum\@tempcnta=\@tempcntb%
     \else \def\@citea{--}\fi
    \advance\@tempcnta\m@ne\the\@tempcnta\@citea\the\@tempcntb}\fi\fi}
\makeatother
\begin{document}
%
\vspace*{-1.0cm}
\begin{center}
{\Large\bf Scaling Calculation of Isoscalar Giant Resonances
in Relativistic Thomas--Fermi Theory}
\\[2.0cm]
S.K. Patra\footnote{Present address:
           {\it Institute of Physics, Sachivalaya Marg,
                Bhubaneswar-{\sl 751 005}, India}},
X. Vi\~nas, M. Centelles and M. Del Estal \\[2mm]
{\it Departament d'Estructura i Constituents de la Mat\`eria,
     Facultat de F\'{\i}sica,
\\
     Universitat de Barcelona,
     Diagonal {\sl 647}, E-{\sl 08028} Barcelona, Spain}
\end{center}
%
\vspace*{2.0cm}
\begin{abstract}
We derive analytical expressions for the excitation energy of the
isoscalar giant monopole and quadrupole resonances in finite nuclei,
by using the scaling method and the extended Thomas--Fermi approach to
relativistic mean field theory. We study the ability of several
non-linear $\sigma-\omega$ parameter sets of common use in reproducing
the experimental data. For monopole oscillations the calculations
agree better with experiment when the nuclear matter incompressibility
of the relativistic interaction lies in the range 220--260 MeV\@. The
breathing-mode energies of the scaling method compare satisfactorily
with those obtained in relativistic RPA and time-dependent mean field
calculations. For quadrupole oscillations all the analyzed non-linear
parameter sets reproduce the empirical trends reasonably well.
\end{abstract}

\mbox{}



\pagebreak

%
\section{Introduction}

The relativistic mean field (RMF) approach to Quantum Hadrodynamics
\cite{serot86} has become a very useful tool for describing
ground-state properties of nuclei along the periodic table. The
simplest model, the linear $\sigma-\omega$ model of Walecka
\cite{Wa74}, describes the nuclear force in terms of the exchange of
$\sigma$ and $\omega$ mesons. It is known that the value of the
nuclear matter incompressibility is unreasonably high in this linear
model ($K_\infty \sim 550$ MeV), which is a serious drawback for a
precise description of some properties of finite nuclei and of
collective excitations such as the breathing mode (isoscalar giant
monopole resonance). The problem can be cured by introducing cubic and
quartic self-interactions of the $\sigma$ meson \cite{boguta77}, which
in particular have the effect of lowering the incompressibility, and
the model can be refined by adding an isovector $\rho$ meson. Current
non-linear parameter sets, such as the NL3 parametrization
\cite{lalaz97}, give ground-state binding energies and densities in
very good agreement with the experimental data, not only for magic
nuclei but also for deformed nuclei as well as for nuclei far from the
stability line.

The RMF model has also been applied to describe dynamical collective
motions in nuclei. The basic theory of vibrational states in nuclei,
the random-phase approximation (RPA) \cite{bertsch75,ring80}, has been
generalized to the relativistic domain (RRPA)
\cite{muillier89,dawson90,piekarewicz01} and it has been used in
calculations of isoscalar giant resonances, to obtain response
functions and mean energies for several magic nuclei. Small-amplitude
collective motions such as the isovector dipole oscillation and the
isoscalar and isovector quadrupole oscillations \cite{vretenar95}, as
well as the isoscalar and isovector monopole oscillations
\cite{vretenar97}, have been studied in the time-dependent RMF
approach. Another approach is based on constrained RMF calculations.
It has been applied to obtain breathing-mode energies and
incompressibilities in the linear \cite{maruyama89,boersma91} and
non-linear \cite{stoi94,stoitsov94} $\sigma-\omega$ models. The
generator coordinate method, with generating functions that are
solutions of constrained RMF calculations, has been employed to
compute excitation energies and transition densities of giant monopole
states \cite{vretenar97,stoi94}. Other calculations of breathing-mode
energies in the relativistic framework, see Refs.\
\cite{stoitsov94,eiff94,chossy97}, have relied on the scaling model in
combination with the leptodermous expansion of the finite nucleus
incompressibility derived by Blaizot \cite{blaizot80}.

In the non-relativistic framework it is well established that the RPA
is the small amplitude limit of the time-dependent Hartree--Fock
approach \cite{ring80,bohigas79}. In the relativistic case the RPA
configuration space must include negative energy states from the Dirac
sea in order to reproduce the results of time-dependent RMF or
constrained RMF calculations \cite{piekarewicz01,ma01}.
Paraphrasing the statement, the RRPA corresponds to the small
amplitude limit of the time-dependent RMF theory in the no-sea
approximation when the RRPA includes both positive energy
particle-hole pairs, and pairs formed from the empty Dirac sea states
and the occupied Fermi sea states.

Semiclassical methods in nuclear physics, like the Thomas--Fermi
theory, have proven to be very helpful for dealing with nuclear
properties of global character that vary smoothly with the particle
number $A$ (e.g., binding energies, densities and their moments)
\cite{ring80,dreizler85,brack97,petkov91}. The success of these
methods stems from the fact that the shell corrections (quantal
effects) are small as compared to the smooth part given by the
semiclassical calculation. Semiclassical techniques like nuclear fluid
dynamics \cite{eckart81} and the extended Thomas--Fermi method
\cite{brack85,centelles90,gleissl90} have been applied to study giant
resonances in non-relativistic models. In the relativistic context,
the nuclear fluid dynamics approach has been utilized, e.g., in Refs.\
\cite{walecka80,furnstahl85,providencia}. The authors of
Refs.\ \cite{nishizaki87,zhu91} resorted to a local Lorentz boost and
the scaling method to study isoscalar giant monopole and quadrupole
states in the linear $\sigma-\omega$ model. The investigations were
carried out for nuclear matter (where a Thomas--Fermi approximation
is exact) \cite{nishizaki87} and for symmetric, uncharged finite
nuclei \cite{zhu91} whose densities were solved in the
relativistic Thomas--Fermi (RTF) approximation.

The relativistic extended Thomas--Fermi (RETF) method
\cite{baranco90,centelles93a,speicher92} is a refinement of
the RTF method, which incorporates gradient corrections of order
$\hbar^2$ to the pure RTF approximation. It was derived only a few
years ago and it has since been applied in calculations of
ground-state binding energies and radii of finite nuclei
\cite{centelles93a,centelles92,speicher93} and in investigations of
nuclear surface properties
\cite{centelles93a,speicher93,centelles93,centelles98}. In the present
work we shall use the RTF and RETF approaches to calculate the
excitation energies of the isoscalar giant monopole and quadrupole
resonances in spherical nuclei. This will be done by means of the
scaling method, within the framework of the non-linear $\sigma-\omega$
model and the RMF theory. We shall also perform constrained
calculations for the monopole state.

Recently, the basic theory derived in the RTF approach has been
applied to discuss the virial theorem and to study the breathing-mode
energy within the RMF theory \cite{patra01}. In the present
contribution we analyze our self-consistent method in depth. To our
knowledge, for realistic non-linear parameter sets of the RMF theory,
these are the first calculations of isoscalar giant resonances in
finite nuclei carried out with the scaling method which are fully
self-consistent (i.e., we do not make use of a leptodermous expansion
as in previous scaling approaches \cite{stoitsov94,eiff94,chossy97}).
Owing to the meson-exchange nature of the relativistic model one has
to deal with finite range forces, which renders the scaling method
more involved than, e.g., for non-relativistic zero-range Skyrme
forces. Moreover, in contrast to the non-relativistic situation, there
exist two different densities, namely the baryon and the scalar
density, in accordance with the fact that one has two types of fields,
the vector and the scalar field.

The article is organized as follows. After the introductory remarks,
we collect the basic expressions of the energy density and the
variational equations of the RTF and RETF models in Section 2. The
third and fourth sections are devoted to the derivation of the
equations and the discussion of the numerical applications for the
giant monopole and quadrupole resonances, respectively. The
conclusions are laid in the last section. Some technicalities and a
derivation of the virial theorem for the relativistic model are given
in the appendices.

\pagebreak

\section{Energy density and variational equations}
\label{energy}

The mean field Hartree energy density of a finite nucleus in the
non-linear $\sigma-\omega$ model reads
\cite{serot86,Wa74,boguta77}
\beq
{\cal H} = \sum_i \varphi_i^{\dagger}
\left[ - i \vect{\alpha} \cdot \vect{\nabla} +
\beta m^* + g\dr{v} V + \frac{1}{2} g_\rho R \tau_3
+ \frac{1}{2} e {\cal A} (1+\tau_3) \right]
 \varphi_i + {\cal H}\dr{f} .
\label{eqFN1}\eeq
The relativistic effective mass (or Dirac mass) is defined by $m^* = m
- g\dr{s}\phi$, $\tau_3$ is the third component of the isospin
operator, and the subindex $i$ runs over the occupied states
$\varphi_i$ of the positive energy spectrum. ${\cal H}\dr{f}$
stands for the free contribution of the meson fields $\phi$, $V$ and
$R$ associated with the $\sigma$, $\omega$ and $\rho$ mesons,
respectively, and of the Coulomb field $\cal A$:
\beqa
{\cal H}\dr{f} & = &
\frac{1}{2} \left[ (\vect{\nabla}\phi)^2 + m\dr{s}^2 \phi^2 \right]
+ \frac{1}{3} b \phi^3 + \frac{1}{4} c \phi^4 
- \frac{1}{2} \left[ (\vect{\nabla} V)^2 + m\dr{v}^2 V^2 \right]
\nonumber \\[3mm]
 & & \mbox{}
- \frac{1}{2} \left[ (\vect{\nabla} R)^2 + m_\rho^2 R^2 \right]
- \frac{1}{2} \left(\vect{\nabla}  {\cal A}\right)^2 .
\label{eqFN1c}
\eeqa
It is understood that the densities and fields are local quantities
that depend on position, even if we do not make it explicit in most
of our expressions. Units are $\hbar= c = 1$.

The semiclassical representation of the energy density \req{eqFN1} has
a similar structure, except that the nucleon variables are the neutron
and proton densities ($\rho\dr{n}$ and $\rho\dr{p}$) instead of the
wave functions. In the RETF approach it reads
\cite{centelles93a,centelles92,speicher93,centelles93,centelles98}
\beq
{\cal H} = {\cal E} + g\dr{v} V \rho 
+ g_\rho R \rho_3 + e {\cal A} \rho\dr{p} + {\cal H}\dr{f} ,
\label{eqFN2}\eeq
where $\rho= \rho\dr{p}+\rho\dr{n}$ is the baryon density, $\rho_3=
{\textstyle{1\over2}} (\rho\dr{p}-\rho\dr{n})$ is the isovector
density, and the nucleon energy density ${\cal E}$ is written as 
${\cal E} = {\cal E}_0 + {\cal E}_2$ with
\beq
{\cal E}_0 =
\sum_{q} \frac{1}{8\pi^2} \left[k_{{\rm F}q}\epsilon^{3}_{{\rm F}q}
+k^{3}_{{\rm F}q}\epsilon_{{\rm F}q}
-{m^*}^{4}\ln\frac{k_{{\rm F}q}+\epsilon_{{\rm F}q}}{m^*}\right]
\label{eqFN2b}
\eeq
and
\beq
{\cal E}_2 =  \sum_{q} \left[ B_{1q}(k_{{\rm F}q}, m^*)
 (\vect{\nabla} \rho_{q})^2
+ B_{2q} (k_{{\rm F}q}, m^*)\left( \vect{\nabla}
 \rho_{q} \cdot \vect{\nabla} m^* \right)
+ B_{3q}(k_{{\rm F}q}, m^*)
  (\vect{\nabla} m^*)^2 \right] .
\label{eqFN2c}\eeq
For each kind of nucleon ($q= {\rm n}, {\rm p}$), the local Fermi
momentum $k_{{\rm F}q}$ and $\epsilon_{{\rm F}q}$ are defined by
\beq
k_{{\rm F}q}= (3\pi^2 \rho_{q})^{1/3} , \qquad
\epsilon_{{\rm F}q}=\sqrt{k^2_{{\rm F}q}+{m^*}^2} .
\eeq
The coefficients $B_{iq}$ are the following functions of $k_{{\rm
F}q}$ and $m^*$ \cite{centelles93a,centelles98}:
\beqa
B_{1q} & = & \frac{\pi^2}{24 k_{{\rm F}q}^3
\epsilon_{{\rm F}q}^2}
\left( \epsilon_{{\rm F}q} + 2k_{{\rm F}q} \ln \frac{k_{{\rm F}q} +
\epsilon_{{\rm F}q}}{m^*} \right) ,
\nonumber \\[3mm]
B_{2q} & = &
\frac{m^*}{6 k_{{\rm F}q} \epsilon_{{\rm F}q}^2}
\ln \frac{k_{{\rm F}q} + \epsilon_{{\rm F}q}}{m^*} ,
\nonumber\\[3mm]
B_{3q} & = &
\frac{k_{{\rm F}q}^2}{24 \pi^2 \epsilon_{{\rm F}q}^2}
\left[\frac {\epsilon_{{\rm F}q}}{k_{{\rm F}q}}
- \left( 2 + \frac{\epsilon_{{\rm F}q}^2}{k_{{\rm F}q}^2} \right)
\ln \frac{k_{{\rm F}q} + \epsilon_{{\rm F}q}}{m^*} \right] .
\label{eqA11} \eeqa
The RTF approximation is obtained by neglecting ${\cal E}_2$ in Eq.\
\req{eqFN2}. The gradients contained in ${\cal E}_2$ arise from the
RETF corrections of order $\hbar^2$ to the functional ${\cal E}_0$.
Naturally, these corrections are more important in the nuclear
surface region where the densities and the fields change more rapidly.

The semiclassical ground-state densities and meson fields are obtained
by solving the Euler--Lagrange equations $\delta {\cal H}/\delta
\rho_q = \mu_q$ (with $\mu_q$ being the chemical potential) coupled to
the field equations
\beqa
(\Delta- m\dr{s}^2)\phi & = & -g\dr{s} \rho\dr{s} +b\phi^2 +c\phi^3 ,
\label{eqFN4}  \\[3mm]
   (\Delta - m\dr{v}^2) V  & = &   -g\dr{v} \rho ,
\label{eqFN5}  \\[3mm]
   (\Delta - m_\rho^2)  R  & = &  - g_\rho \rho_3 ,
\label{eqFN6}  \\[3mm]
 \Delta {\cal A}   & = &     -e \rho\dr{p} .
\label{eqFN7}
\eeqa
The semiclassical scalar density in \req{eqFN4} is given by
\beqa
\rho\dr{s} & = & \frac{\delta {\cal E}_0}{\delta m^*}
               + \frac{\delta {\cal E}_2}{\delta m^*}
 = \rho\dr{s0} + \rho\dr{s2}
\nonumber \\[3mm]
& = &
 \sum_{q} \frac{m^*}{2\pi^2}\left[k_{{\rm F}q}\epsilon_{{\rm
F}q}-{m^*}^2 \ln\frac{k_{{\rm F}q}
+\epsilon_{{\rm F}q}} {m^*}\right]
\nonumber \\[3mm]
& &
- \sum_q \left[ B_{2q}\Delta\rho_{q}+2B_{3q}\Delta m^*
+\left(\frac{\partial B_{2q}}
{\partial \rho_{q}}-\frac{\partial B_{1q}}{\partial m^*}\right)
(\vect{\nabla} \rho_q)^2 \right.
\nonumber \\[3mm]
& & \left. \mbox{}
+2 \frac{\partial B_{3q}}{\partial \rho_{q}}
(\vect{\nabla} \rho_{q} \cdot \vect{\nabla} m^*)
 + \frac{\partial B_{3q}}{\partial m^*}
(\vect{\nabla} m^* )^2 \right] .
\label{eqFN8}
\eeqa
Parenthetically, we would like to mention that the densities $\rho$,
$\rho_{\rm s}$ and $\cal E$ above are the semiclassical counterparts
of the quantal densities $\rho= \sum_i \varphi_i^{\dagger} \varphi_i$,
$\rho_{\rm s}= \sum_i \varphi_i^\dagger \beta \varphi_i$ and ${\cal
E}= \sum_i \varphi_i^{\dagger} \left[ - i \vect{\alpha} \cdot
\vect{\nabla} + \beta m^* \right] \varphi_i$.

Since the energy density $\cal H$ is to be integrated over the space
to compute the total energy, the field equations
\req{eqFN4}--\req{eqFN7} can be used to rewrite ${\cal H}\dr{f}$,
e.g., by transforming $[(\vect{\nabla}V)^2 + m\dr{v}^2 V^2]$ into $V
(-\Delta + m\dr{v}^2) V = g\dr{v} V \rho$ (valid, of course, under an
integral sign). This way, on defining an effective scalar density by
\beq 
g\dr{s} \rho\ur{eff}\dr{s} = g\dr{s} \rho\dr{s} 
-b \phi^2 - c \phi^3 , 
\label{eqFN8b}\eeq
${\cal H}$ can be recast as 
\beq  
{\cal H} = {\cal E} + \frac{1}{2}g\dr{s}\phi \rho\ur{eff}\dr{s}
+ \frac{1}{3}b\phi^3+\frac{1}{4} c \phi^4 
+\frac{1}{2} g\dr{v}  V \rho +\frac{1}{2} g_\rho R \rho_3
+\frac{1}{2} e {\cal A} \rho\dr{p} .
\label{eqFN8c}\eeq
This form of ${\cal H}$ will be more convenient for facilitating the
calculations to be presented below.

\pagebreak

\section{Giant monopole resonance}

As far as the giant resonances are dynamical processes one must first
describe the nucleus from a moving frame \cite{nishizaki87,zhu91}.
This is a rather technical matter for our present purposes and it is
left for Appendix~A, where we derive the expression of the energy of a
nucleus within the relativistic model in a frame moving with velocity
$-\vect{v}$. After performing the scaling of the energy in this frame,
one obtains general expressions for the two main ingredients required
for the calculation of the excitation energy of the giant resonance,
namely, the restoring force and the mass or inertia parameter (Eqs.\
\req{eqFNA15} and \req{eqFNA16}, respectively).

The present section proceeds as follows. We begin by introducing our
scaling approach for the monopole vibration. Next we obtain analytical
expressions for the restoring force and the mass parameter of the
monopole state. The calculational details of the derivatives of the
meson fields with respect to the collective coordinate of the monopole
vibration are reserved for Appendix B, where we also discuss the
virial theorem (stationarity of the scaled energy) for the
relativistic model. Next in the section we give a brief summary of the
constrained approach to the breathing mode, for comparison with the
scaling approach. The section closes with the discussion of the
results of our numerical calculations.

\subsection{Scaling}

Denoting by $\lambda$ the collective coordinate associated with the
monopole vibration, a normalized scaled version of the baryon density
is 
\beq
 \rho_\lambda (\vect{r}) = \lambda^3 \rho(\lambda\vect{r}) .
\label{eqFN10}\eeq
Accordingly, the local Fermi momentum changes as
\beq
 k_{{\rm F}q\lambda} (\vect{r})  = 
 [ 3\pi^2\rho_{q\lambda}(\vect{r}) ]^{1/3}=
\lambda k_{{\rm F}q} (\lambda\vect{r}) .
\label{eqFN11}\eeq
The meson fields $\phi$, $V$ and $R$ and the Coulomb field ${\cal A}$
are also modified by the scaling due to the self-consistent equations
\req{eqFN4}--\req{eqFN7}, which will relate the scaled fields to the
scaled densities. Unfortunately, the meson fields do not scale
according to simple laws like \req{eqFN10} and \req{eqFN11} because of
the finite range of the meson interactions. This is most apparent for
the scalar field $\phi$, since the scalar density in the source term
of Eq.\ (\ref{eqFN4}) transforms not only due to the scaling of
$k_{{\rm F}q}$ but also of $\phi$ itself (or $m^*$), see Eq.\
(\ref{eqFN8}) for $\rho_{\rm s}$. For reasons that will become clear
immediately, we shall write the scaled effective mass
$m^*_\lambda(\vect{r})= m - g_{\rm s} \phi_\lambda(\vect{r})$ in the
form 
\begin{equation}
 m^*_\lambda(\vect{r}) \equiv \lambda {\tilde m}^* (\lambda\vect{r}) .
\label{eqFN13} \end{equation}
Note that the quantity ${\tilde m}^*$ carries an implicit dependence
on $\lambda$ apart from the parametric dependence on
$\lambda\vect{r}$. 

On account of Eqs.\ \req{eqFN11} and \req{eqFN13} the scaled
semiclassical energy density ${\cal E}_\lambda= {\cal E}_{0\lambda} +
{\cal E}_{2\lambda}$ and scalar density $\rho\dr{s\lambda} =
\rho\dr{s0\lambda}+\rho\dr{s2\lambda}$ read
\beqa
{\cal E}_\lambda(\vect{r}) & = & 
\lambda^4 {\cal E}_0 [ k_{{\rm F}q} (\lambda\vect{r}),
                         {\tilde m}^* (\lambda\vect{r}) ] +
\lambda^4 {\cal E}_2 [ k_{{\rm F}q} (\lambda\vect{r}),
                         {\tilde m}^* (\lambda\vect{r}) ] \equiv
\lambda^4 {\tilde{\cal E}} (\lambda\vect{r}) ,
\label{eqFN15} \\[3mm]
\rho\dr{s\lambda} (\vect{r}) & = & 
\lambda^3 \rho\dr{s0} [ k_{{\rm F}q} (\lambda\vect{r}),
                         {\tilde m}^* (\lambda\vect{r}) ] +
\lambda^3 \rho\dr{s2} [ k_{{\rm F}q} (\lambda\vect{r}),
                         {\tilde m}^* (\lambda\vect{r}) ] \equiv
\lambda^3 {\tilde\rho}\dr{s} (\lambda\vect{r}) .
\label{eqFN16}\eeqa
The tilded quantities ${\tilde{\cal E}}$ and ${\tilde\rho}\dr{s}$ are
given by Eqs.\ \req{eqFN2b}, \req{eqFN2c} and \req{eqFN8} after
replacing $m^*$ by ${\tilde m}^*$. Note the usefulness of \req{eqFN13}
to be able to write \req{eqFN15} and \req{eqFN16} in this compact
form. For the scaled total energy density ${\cal H_\lambda}$ we obtain
\beq
{\cal H}_\lambda  = \lambda^3 \bigg[ \lambda {\tilde{\cal E}}
+\frac{1}{2}g\dr{s} \phi_\lambda {\tilde\rho}\ur{eff}\dr{s}
+\frac{1}{3} \frac{b}{\lambda^3}\phi_\lambda^3
+\frac{1}{4} \frac{c}{\lambda^3} \phi_\lambda^4 
+\frac{1}{2} g\dr{v}  V_\lambda \rho
+\frac{1}{2}g_\rho R_\lambda \rho_3 
+\frac{1}{2} e {\cal A}_\lambda\rho\dr{p} \bigg],
\label{eqFN17}\eeq
with the definition
\beq
g\dr{s}{\tilde\rho}\ur{eff}\dr{s} = g\dr{s} {\tilde\rho}\dr{s}
-\frac{b}{\lambda^3} \phi_\lambda^2 - \frac{c}{\lambda^3}
\phi_\lambda^3.
\label{eqFN18}\eeq
Observe that the same expression is valid regardless of performing the
calculations in the RETF model or in the RTF model, as the corrections
of order $\hbar^2$ (${\cal E}_2$ and $\rho\dr{s2}$) scale in the same
manner as the the Thomas--Fermi terms (${\cal E}_0$ and
$\rho\dr{s0}$).

\subsection{Restoring force}
\label{restoring}

The restoring force $C\dr{M}$ of the monopole vibration is given by
the second derivative of the scaled energy with respect to the
collective coordinate $\lambda$, calculated at $\lambda=1$ (Appendix
A). The first derivative of the scaled energy is
\beqa
\frac{\partial}{\partial \lambda} \int
 d (\lambda \vect{r}) 
\frac{ {\cal H}_\lambda (\vect{r})}{\lambda^3} & = &
 \int d (\lambda \vect{r}) \left[ {\tilde{\cal E}}
 - {\tilde m}^* {\tilde\rho}\dr{s} 
-\frac{1}{2}g\dr{s}{\tilde\rho}\ur{eff}\dr{s}
\frac{\partial \phi_\lambda}{\partial \lambda} 
+\frac{1}{2}g\dr{s} \phi_\lambda  
\frac{\partial {\tilde\rho}\ur{eff}\dr{s}}{\partial \lambda}
-\frac{b}{\lambda^4}\phi_\lambda^3 \right.
\nonumber \\[3mm]
& & \left. \mbox{}
-\frac{3}{4} \frac{c}{\lambda^4}\phi_\lambda^4 
+\frac{1}{2}g\dr{v}\rho\frac{\partial  V_\lambda}{\partial \lambda}
+\frac{1}{2}g_\rho \rho_3 \frac{\partial R_\lambda}
{\partial \lambda}+\frac{1}{2}e\rho\dr{p} \frac{\partial
{\cal A}_\lambda} {\partial \lambda} \right] .
\label{eqFN19}
\eeqa
Here we have used $\partial {\tilde{\cal E}}/\partial \lambda=
{\tilde\rho}_{\rm s} \, \partial {\tilde m}^*/\partial \lambda$ (as
${\tilde\rho}_{\rm s}= \delta {\tilde{\cal E}}/\delta {\tilde m}^*$)
and 
\beq
\frac{\partial m^*_\lambda}{\partial \lambda} =
{\tilde m}^* + \lambda \frac{\partial {\tilde m}^*}
{\partial \lambda}=-g\dr{s} \frac{\partial \phi_\lambda}{\partial
\lambda} ,
\label{eqFN20} \eeq
from the definition \req{eqFN13} of ${\tilde m}^*$. Differentiating
again with respect to $\lambda$ and then setting $\lambda=1$ we have
\beqa  & &
 C\dr{M} = \left[ \frac{\partial^2}{\partial \lambda^2} \int
 d (\lambda \vect{r}) 
\frac{ {\cal H}_\lambda (\vect{r})}{\lambda^3} \right]_{\lambda=1}  = 
\nonumber \\[3mm]
 & &
\int d \vect{r} \left[ - {\tilde m}^*
\frac{\partial {\tilde\rho}\dr{s}}{\partial \lambda}
-\frac{1}{2}g\dr{s}{\tilde\rho}\ur{eff}\dr{s}
\frac{\partial^2\phi_\lambda}{\partial \lambda^2}
+\frac{1}{2}g\dr{s} \phi_\lambda
\frac{\partial^2 {\tilde\rho}\ur{eff}\dr{s}} {\partial\lambda^2} 
+ 4 \frac{b}{\lambda^5}\phi^3_\lambda
+ 3 \frac{c}{\lambda^5}\phi^4_\lambda
\right.
\nonumber \\[3mm]
 & &
\left. \mbox{} 
- \frac{3}{\lambda^4} ( b \phi^2_{\lambda} + c \phi^3_{\lambda} )
 \frac{\partial \phi_\lambda} {\partial \lambda}
+\frac{1}{2}g\dr{v}\rho \frac{\partial^2  V_\lambda}
{\partial \lambda^2} +\frac{1}{2}g_\rho \rho_3
\frac{\partial^2  R_\lambda}{\partial \lambda^2}
+\frac{1}{2}e\rho\dr{p}
\frac{\partial^2  {\cal A}_\lambda}{\partial \lambda^2}
\right]_{\lambda=1} . 
\label{eqFN27}\eeqa

The calculation of the derivatives of the scaled meson fields with
respect to $\lambda$ is illustrated in Appendix B\@. There we also
work out Eq.\ \req{eqFN19} at $\lambda=1$, which leads to the virial
theorem (stationarity condition of the energy) for the relativistic
mean field model. Following the technique outlined in Appendix B we
find 
\beq
\left. \frac{\partial^2  V_\lambda(\vect{r})}{\partial
\lambda^2}\right|_{\lambda=1}
 =  \int d \vect{r}' \rho(\vect{r}')
\left[ 2s \frac{d{\cal V}_\omega(s)}{d s}
+s^2 \frac{d^2{\cal V}_\omega(s)}{d s^2} \right] ,
\label{eqFN28}\eeq
with
\beq
{\cal V}_\omega (s) = \frac{g\dr{v}}{4\pi}
\frac{e^{-m\dr{v} s}}{s} ,
\qquad
s=|\vect{r}-\vect{r}'| .
\label{eqFN22}\eeq
An equivalent expression holds for the field $R_\lambda$. As is well
known, the second derivative of the scaled Coulomb field ${\cal
A}_\lambda$ vanishes \cite{bohigas79} (you only have to evaluate
\req{eqFN28} for a zero meson mass). For the scalar field one gets
a lengthier expression owing to the extra implicit dependence of
${\tilde\rho}\ur{eff}\dr{s}$ on $\lambda$:
\beqa
\left. \frac{\partial^2 \phi_\lambda(\vect{r})}
{\partial \lambda^2}\right|_{\lambda=1}
& = & 
\int d\vect{r}'\rho\dr{s}\ur{eff}(\vect{r}')
\left[2s \frac{d{\cal V}_\sigma(s)}{d s}+s^2 \frac{d^2{\cal
V}_\sigma(s)}{d s^2} \right] 
\nonumber \\[3mm]
& &
- \int d \vect{r}' \left[
2 s \frac{d {\cal V}_\sigma(s)}{d s}
\frac{\partial{\tilde\rho}\ur{eff}\dr{s} (\lambda\vect{r}')}
{\partial\lambda} - {\cal V}_\sigma (s)
\frac{\partial^2 {\tilde\rho}\ur{eff}\dr{s} (\lambda\vect{r}')}
{\partial\lambda^2} \right]_{\lambda=1} .
\label{eqFN29}\eeqa

Inserting these results into Eq.\ \req{eqFN27} for $C\dr{M}$ we end up
with 
\beqa
C\dr{M}  & = & 
\int d \vect{r} \left\{ \left[ -{\tilde m^*}
\frac{\partial {\tilde\rho}\dr{s}}{\partial \lambda}
+ g\dr{s} \frac{\partial{\tilde\rho}\ur{eff}\dr{s}}{\partial\lambda}
\int d\vect{r}'\rho\dr{s}\ur{eff}(\vect{r}')
 s \frac{d {\cal V}_\sigma}{d s} 
- 3(b\phi^2+c\phi^3) \frac{\partial \phi_\lambda}
{\partial \lambda}\right]_{\lambda=1}  \right.
\nonumber \\[3mm]
& & \mbox{}
-\frac{1}{2}g\dr{s} \rho\dr{s}\ur{eff}
\int d\vect{r}' \rho\dr{s}\ur{eff}(\vect{r}')
\left[2s \frac{d{\cal V}_\sigma}{d s}
+s^2 \frac{d^2{\cal V}_\sigma}{d s^2}\right]
+4b\phi^3+3c\phi^4
\nonumber \\[3mm]
& & \mbox{} 
+ \frac{1}{2}g\dr{v} {\rho}\int d\vect{r}' \rho(\vect{r}')
\left[2s \frac{d{\cal V}_\omega}{d s}
+s^2 \frac{d^2{\cal V}_\omega}{d s^2}\right]
\nonumber \\[3mm]
& & \left. \mbox{}
+\frac{1}{2}g_\rho \rho_3 \int d\vect{r}' \rho_3(\vect{r}')
\left[2s \frac{d{\cal V}\dr{\rho}}{d s}
+s^2 \frac{d^2{\cal V}\dr{\rho}}{d s^2} \right] \right\} ,
\label{eqFN30}\eeqa
After some algebra it is possible to recast the restoring force of
the monopole state as
\begin{eqnarray} 
 C_{\rm M} & = &
\int d \vect{r} \left[ 
- m \frac{\partial {\tilde\rho}_{\rm s}}{\partial \lambda} 
+ 3 \left( m_{\rm s}^2 \phi^2 + \frac{1}{3} b \phi^3 
- m_{\rm v}^2 V^2 - m_\rho^2 R^2 \right)  \right.
\nonumber \\[1.5mm]
& & \left. \mbox{}
- (2 m_{\rm s}^2 \phi + b\phi^2) 
\frac{\partial \phi_\lambda}{\partial\lambda}
+ 2 m_{\rm v}^2 V
\frac{\partial V_\lambda}{\partial\lambda} 
+ 2 m_\rho^2 R
\frac{\partial R_\lambda}{\partial\lambda} \right]_{\lambda=1} .
\label{eqFN30b}\end{eqnarray}
Note that as in the case of ${\cal H}_\lambda$ the same expression
holds in both the RTF and RETF models. In each model one just has to
compute $C\dr{M}$ with the ground-state densities and fields obtained
from the solution of the corresponding variational equations (which
are modified by the inclusion of the corrections ${\cal E}_2$ and
$\rho\dr{s2}$). The derivatives with respect to $\lambda$ entering
Eq.\ \req{eqFN30b} can be calculated as indicated in Appendix B, Eqs.\
(\ref{eqFN23})--(\ref{eqFN245}). We have found $C_{\rm M}$ to take
positive values for all of the (linear and non-linear) parameter sets
we have used in the RTF and RETF calculations. A large part of the
final value of $C_{\rm M}$ (usually far more than a half) is due to
the contribution of the term $-m \partial {\tilde\rho}_{\rm
s}/\partial \lambda|_{\lambda=1}$.

Evaluation of the integrand of Eq.\ \req{eqFN30b} in the limit of
symmetric infinite nuclear matter gives the result
\beq
K_\infty \rho_\infty = 9 \frac{g^2\dr{v}}{m\dr{v}^2} \rho_\infty^2
+ 3 \frac{k^2\dr{F,\infty}}{\epsilon\dr{F,\infty}} \rho_\infty
+3  \frac{m^*_\infty}{\epsilon\dr{F,\infty}} \rho_\infty
\left. \frac{\partial {\tilde m}^*_\infty}
{\partial \lambda}\right|_{\lambda=1} .
\label{eqFN31} \eeq
From the field equation for the scaled scalar field in nuclear matter
we have
\beq
\left. \frac{\partial {\tilde m}^*_\infty}
{\partial \lambda}\right|_{\lambda=1} =
- 3 g^2\dr{s}
\frac{m^*_\infty}{\epsilon\dr{F,\infty}} \rho_\infty
\left[ m\dr{s}^2 + 3 g^2\dr{s}
\left( \frac{\rho\dr{s,\infty}}{m^*_\infty}
-\frac{\rho_\infty}{\epsilon\dr{F,\infty}} \right)
+ 2 b \phi_\infty^2 + 3 c \phi_\infty^3 \right]^{-1} ,
\label{eqFN32}\eeq
and, as expected, $K_\infty$ in \req{eqFN31} is seen to coincide with
the expression of the bulk nuclear incompressibility in the
relativistic model \cite{speicher93,centelles93}.

\subsection{Mass parameter}

As explained in Appendix A the mass parameter of the giant resonance
is obtained from the second derivative of the scaled energy in a
moving frame with respect to the collective velocity
$\dot\lambda=d\lambda/dt$, see Eq.\ \req{eqFNA16}. The relation
between the collective velocity $\dot\lambda$ and the velocity
$\vect{v}$ of the moving frame is provided by the continuity equation
\req{eqFN35} for the scaled system. This equation suggests a radial
velocity field of the form $\vect{v}= - \dot\lambda u(r) \vect{r}/r$
up to first order in $\dot\lambda$ for the monopole mode
\cite{maruyama89,boersma91,stoitsov94}. In terms of the displacement
field $u(r)$ the mass parameter \req{eqFNA16} is written as
\beq
B =  \int d\vect{r} u^2(r) {\cal H} ,
\label{eqFN35g}\eeq
while the continuity equation \req{eqFN35} becomes
\beq
\frac{d \rho_\lambda(\vect{r})}{d \lambda} - {\nabla} \cdot 
\left[ \rho_\lambda(\vect{r}) u(r) \frac{\vect{r}}{r} \right] = 0 .
\label{eqFN35h}\eeq
At $\lambda=1$ Eq.\ \req{eqFN35h} determines the displacement field as
\beq
u(r)= \frac{1}{\rho(r)r^2} \int_0^r d r' r'^2 \rho_{\rm T}(r') ,
\label{eqFN35a}\eeq
where 
\beq
\rho_{\rm T}(\vect{r}) = \left. 
\frac{d \rho_\lambda(\vect{r})}{d \lambda} \right|_{\lambda=1}
\label{eqFN35k}\eeq
is the so-called transition density.

For the monopole mode $\rho_\lambda(\vect{r}) = \lambda^3 \rho(\lambda
\vect{r})$ and, thus,  the transition density is given by
\beq
\rho_{\rm T}(r) = 3\rho(r) + r \frac{d\rho(r)}{dr} ,
\label{eqFN35b}\eeq
which is known as the Tassie transition density. Partial integration
of Eq.\ \req{eqFN35a} with \req{eqFN35b} leads to the well-known
result $u(r)=r$ for the displacement field under the scaling
transformation. The mass parameter of the monopole oscillation thus
becomes \cite{nishizaki87,zhu91}
\beq
B\dr{M} =  \int d\vect{r} r^2 {\cal H} .
\label{eqFN36}\eeq
Finally one calculates the excitation energy of the monopole state in
the scaling model as
\beq
{\bar E}\dr{M}\ur{s} = \sqrt{ \frac{C\dr{M}}{B\dr{M}} } .
\label{eqFN36b}\eeq

Let us mention in passing that in the non-relativistic approach the
mass parameter is derived as $B\dr{M}^{\rm nr} = \int d\vect{r} r^2 m
\rho$, with $m$ being the nucleon mass. Actually, provided that
$B\dr{M}$ is replaced by $B\dr{M}^{\rm nr}$ in \req{eqFN36b}, the
scaling energy of the monopole vibration can be formally expressed as
in the non-relativistic sum-rule approach. That is, ${\bar
E}\dr{M}\ur{s}= \sqrt{m_3/m_1}$, where the moment $m_1$ is the energy
weighted sum rule
\beq  m_1= 
 \frac{2}{m} A \langle r^2 \rangle = \frac{2}{m^2} B\dr{M}^{\rm nr} ,
\label{eqsumr1}\eeq
and $m_3$ is the plus three energy moment related to the second
derivative of the scaled energy \cite{bohigas79}.

\subsection{Constrained calculation}
\label{constrained}

The giant monopole resonance can also be studied by performing a
constrained calculation. In the semiclassical context one has to
minimize the constrained functional
\beq
 \int d \vect{r} [ {\cal H} - \eta r^2 \rho ] =
 E(\eta)  - \eta \int d \vect{r} r^2 \rho
\label{eqFN56}\eeq
with respect to arbitrary variations of the proton and neutron
densities and of the meson fields. The densities, fields and energy
obtained from the solution of the variational equations associated to
\req{eqFN56}, now depend on the value of the parameter $\eta$. The
nuclear r.m.s.\ radius is calculated as
\beq
 R_\eta = \left[\frac{1}{A}\int d \vect{r} r^2 \rho \right]^{1/2} ,
\label{eqFN57}\eeq
where $A$ is the mass number of the nucleus. 

The constrained energy $E(\eta)$ has a minimum at $\eta=0$ which
corresponds to the ground-state r.m.s.\ radius $R_0$. Following Refs.\
\cite{maruyama89,boersma91,stoi94,stoitsov94} one expands $E(\eta)$ in
a harmonic approximation about $R_0$ to obtain the constrained
incompressibility of the finite nucleus as
\beq
K\ur{c}\dr{A} = \frac{1}{A} R_0^2 \left.
\frac{\partial^2 E(\eta)}{\partial R_\eta^2}\right|_{\eta=0} .
\label{eqFN60}\eeq
In the constrained model the displacement field is also given by Eq.\
\req{eqFN35a}; now with a transition density $\rho_{\rm T}(r) = d
\rho(r,\eta) / d s |_{\eta=0}$, where $s=R_\eta/R_0 - 1$ denotes the
collective variable of the constrained monopole oscillation
\cite{maruyama89,boersma91,stoitsov94}. The frequency of
the constrained isoscalar monopole vibration is computed as
\beq
{\bar E}\dr{M}\ur{c} = \sqrt{\frac{A K\ur{c}\dr{A}}{B\dr{M}\ur{c}}} .
\label{eqFN60b}\eeq
Again, we may notice that if in this equation the inertia parameter is
replaced by its non-relativistic limit $B\dr{M}^{\rm nr}$, the energy
of the constrained monopole vibration can be nominally written in
terms of sum rules. In the present case one hase ${\bar
E}\dr{M}\ur{c}= \sqrt{m_1/m_{-1}}$, with $m_{-1}$ being the inverse
energy-weighted sum rule
\beq
 m_{-1}= - \frac{1}{2} A 
 \left. \frac{\partial R_{\eta}^2}{\partial \eta} \right|_{\eta=0} 
= \frac{1}{2} 
\left. \frac{\partial^2 E(\eta)}{\partial \eta^2} \right|_{\eta=0} .
\label{eqsumr2}\eeq

\subsection{Numerical results}

Our RTF and RETF results for the excitation energy of the isoscalar
giant monopole resonance (GMR) obtained in the scaling approach are
displayed in Table 1, together with the empirical estimate $E\dr{x}
\sim 80/A^{1/3}$ MeV \cite{bertrand76}. We have considered the nuclei
$^{40}$Ca, $^{90}$Zr, $^{116}$Sn, $^{144}$Sm and $^{208}$Pb for which
recent experimental data on the GMR are available \cite{young99}, in
addition to $^{16}$O and $^{48}$Ca. We have employed the non-linear
parameter sets NL1 (incompressibility $K_\infty=211$ MeV, effective
mass in nuclear matter $m^*_\infty/m=0.57$) \cite{reinhard86}, NL3
\cite{lalaz97} ($K_\infty=272$ MeV, $m^*_\infty/m=0.60$), NL-SH
\cite{sharma93} ($K_\infty=355$ MeV, $m^*_\infty/m=0.60$) and NL2
\cite{lee86} ($K_\infty=399$ MeV, $m^*_\infty/m=0.67$). The predictive
power of these parametrizations is well known and some examples can be
found, e.g., in Ref.\ \cite{patra97} and references quoted therein.
The relatively new parameter set NL3 is considered to be the most
successful relativistic effective interaction so far. It is to be
noted that the RMF parameter sets are determined by least-squares fits
to ground-state properties like radii, binding energies and spin-orbit
splittings of a few spherical nuclei. Then, there is no further
adjustment to be made in the parameters of the interaction.

From Table 1 we see that the smaller the mass number, the larger is
the monopole excitation energy. The energy of the GMR increases with
increasing $K_\infty$ in the various parameter sets. For example, in
the RETF calculation the excitation energy for $^{208}$Pb is 12.7 MeV
with NL1, while it is 18.4 MeV with NL2. At first glance the
dependence on $K_\infty$ is roughly linear for each nucleus. We
realize that, overall, the importance of the gradient corrections of
order $\hbar^2$ included in the RETF approach is small for the GMR
energy: the RETF energies differ by less than 10\% from the RTF
energies. The largest deviations between the RETF and RTF results are
found for the lighter systems, where the surface terms are
comparatively more important. If we analyze the relative difference
between the RETF and RTF energies it is seen to decrease with
increasing mass number in all sets, with the sole exception of
$^{16}$O with NL3. For example, the difference goes from $-7\%$ in
$^{16}$O to $-1.6\%$ in $^{208}$Pb with NL1. We observe that the sign
of the correction to the energy of the monopole state due to the
$\hbar^2$ terms depends on the value of $K_\infty$. In the case of NL1
the change of RETF with respect to RTF is negative. For NL3 and NL-SH
the change becomes more and more positive with $K_\infty$ (the
effective mass $m^*_\infty/m$ of these two sets being almost the
same). When we look at NL2 the change is again positive, but smaller
in relative value than for NL-SH owing to the larger effective mass of
NL2, which tends to counterbalance the effect of $K_\infty$.

It is usually recognized that microscopic calculations of the
isoscalar GMR energy in nuclei provide a reliable source of
information on the nuclear matter incompressibility $K_\infty$
\cite{blaizot95,farine97}. The value of $K_\infty$ is an important
ingredient not only for the description of finite nuclei but also for
the study of heavy ion collisions, supernovae and neutron stars. In
practice one has several effective interactions which differ mostly by
their prediction for $K_\infty$, but otherwise reproduce
satisfactorily the experimental data on ground-state properties.
Comparison of the calculated GMR energies with experiment restricts
the range of acceptable values for the nuclear matter
incompressibility of the effective nuclear force. From Table 1 we see
that the empirical law $E\dr{x} \sim 80/A^{1/3}$ MeV roughly lies in
between the predictions of the NL1 and NL3 parameter sets, as expected
from the reasonable value of $K_\infty$ in these interactions. On the
contrary, the NL-SH and NL2 parametrizations have too high a value of
$K_\infty$ and clearly overestimate the empirical curve and the
experimental data for all the considered nuclei. In Figure 1 we have
drawn further RETF results for the excitation energy of the monopole
mode in comparison with the experimental data listed in Ref.\
\cite{shlomo93}, as a function of the number of particles of the
nucleus. (The RTF calculation displays basically the same trend as the
RETF results.) Again, it is clear that the NL-SH and NL2 sets are
unable to describe the experimental values. The experimental points
are roughly enclosed within the predictions of the NL1 and NL3 sets.
For medium mass nuclei NL1 is closer to experiment than NL3, while for
heavier nuclei the experimental energies tend to approach the results
obtained with NL3.

A further inspection of Table 1 shows that no RMF parameter set seems
capable of reproducing the mass-number dependence of the experimental
data over the whole analyzed region, particularly in the lighter
nuclei. One should note, however, that the scaling calculation
provides a prediction for the mean value or centroid of the excitation
energy of the resonance. To establish a relation between the
incompressibility $K_\infty$ and the experimentally measured energies
of the monopole mode the most favourable situation is met in heavy
nuclei, where the strength of the GMR is less fragmented than in
medium and light nuclei \cite{young99,shlomo93}. If we take into
account the excitation energies of $^{116}$Sn, $^{144}$Sm and
$^{208}$Pb, according to Table 1 the nuclear matter incompressibility
$K_\infty$ of the relativistic interaction should lie in the range
$220$--$260$ MeV\@. If we only consider $^{144}$Sm and $^{208}$Pb, as
in Ref.\ \cite{ma01}, then $K_\infty$ would be restricted to the range
$230$--$260$ MeV\@. For comparison, the authors of Refs.\
\cite{vretenar97,ma01} conclude that the value of $K_\infty$ should be
close to 250--270 MeV from their time-dependent RMF and relativistic
RPA calculations. The analysis of the relativistic RPA peak energies
reported in Ref.\ \cite{piekarewicz01} for $^{208}$Pb suggests instead
a range 235--250 MeV for $K_\infty$. On the other hand,
non-relativistic Hartree--Fock plus RPA analyses using Skyrme and
Gogny interactions determine $K_\infty$ to be $215\pm15$ MeV
\cite{blaizot95,farine97}, thus lower than in the RMF model.

The restoring force $C\dr{M}$ of the GMR defines the incompressibility
$K\ur{s}_A$ of the finite nucleus in the scaling model through
$C\dr{M}= A K\ur{s}_A$. In the limit of an arbitrarily large
nucleus $K\ur{s}_A$ should approach the nuclear matter value
$K_\infty$, see Eq.\ \req{eqFN31}. This suggests a linear dependence
of the incompressibility of finite nuclei on the bulk
incompressibility $K_\infty$ of the effective interaction. In Figure 2
we display the value of $K\ur{s}_A$ for the nuclei $^{16}$O,
$^{40}$Ca, $^{116}$Sn and $^{208}$Pb obtained from our RETF
calculation, as a function of $K_\infty$. Apart from the parameter
sets discussed in Table 1, we have employed the sets NL-Z, LZ, L1, L2,
L3 and HS compiled in Table 3 of Ref.\ \cite{reinhard89}, the sets
NLB, NLC and NLD from Ref.\ \cite{furnstahl87}, NL-RA1 from Ref.\
\cite{rashdan01} and L0 from Ref.\ \cite{reinhard86}. The bulk
incompressibility of these sets spans a wide range of values, from
$\sim 175$ to $625$ MeV\@. The sets L0, LZ, L1, L2, L3 and HS
correspond to the linear model without scalar self-interactions. 

The results of Figure 2 show a remarkable linear behaviour of
$K\ur{s}_A$ with the compression modulus $K_\infty$. This is more
true for the heavier systems on the one hand, and for the non-linear
parameter sets on the other hand. The linear sets show a considerable
dispersion, but one should take into account that only L0 and LZ were
optimally adjusted to nuclear ground-state properties and that,
furthermore, L1, L2 and L3 do not include the $\rho$ field. The
incompressibilities of $^{208}$Pb and $^{116}$Sn are nearly the same.
To see a perceptible change one has to go to $^{40}$Ca. In Figure 3 we
have drawn the excitation energy ${\bar E}\dr{M}\ur{s}$ of the
monopole state versus the $K_\infty$ incompressibility. As expected
from the pattern displayed by $K\ur{s}_A$, the monopole excitation
energy increases smoothly with the bulk compression modulus, roughly
as a linear function of the square root of $K_\infty$ (in agreement,
e.g., with Refs.\ \cite{piekarewicz01,blaizot95}). Both the effective
mass at saturation $m^*_\infty$ and the mass of the scalar meson
$m\dr{s}$ play a major role in the determination of the nuclear
structure properties in the RMF theory. The effective mass has a
direct influence on the spin--orbit force and the single-particle
levels, while the scalar mass is related with the range of the
attractive part of the effective nuclear force and thus strongly
affects the nuclear surface. One may wonder whether these two
quantities have some effect on the energy of the breathing mode.
Figure 4 shows that this is not the case, as no evident correlation
seems to exist between ${\bar E}\dr{M}\ur{s}$ and the value of
$m^*_\infty$ or $m\dr{s}$.

From the data represented in Figure 2 we obtain the relations
\beqa
 K\ur{s}_A & = & 0.66 K_\infty - 12   \mbox{ MeV}
 \qquad \mbox{for $^{208}$Pb},
\nonumber\\[0mm]
 K\ur{s}_A & = & 0.65 K_\infty - 9~\, \mbox{ MeV}
 \qquad \mbox{for $^{116}$Sn},
\nonumber\\[0mm]
 K\ur{s}_A & = & 0.57 K_\infty - 7~\, \mbox{ MeV}
 \qquad \mbox{for $^{40}$Ca},
\nonumber\\[0mm]
 K\ur{s}_A & = & 0.45 K_\infty - 8~\, \mbox{ MeV}
 \qquad \mbox{for $^{16}$O} .
\label{eqfit0}\eeqa
The expressions for oxygen and calcium are not as meaningful as for
the heavier nuclei, and we give them mostly for the purpose of
illustration. Though there is a dependence on the mass number, the
slope of the linear fits \req{eqfit0} is visibly smaller than unity.
Moreover, we have obtained a non-vanishing constant term. This is
consistent with the leptodermous expansion of the incompressibility of
a finite nucleus, inspired from the liquid drop formula, in which one
separates the volume, surface, symmetry, Coulomb and higher-order
contributions by writing
\beq
K\ur{s}_A = K_\infty + K\ur{s}_{\rm surf} / A^{1/3} +
K_{\rm sym} (N-Z)^2 / A^2 + K_{\rm Coul} Z^2 / A^{4/3} + \cdots \, .
\label{eqLDM}\eeq 
The total incompressibility $K\ur{s}_A$ receives a sizeable
contribution from the surface term and smaller contributions from the
symmetry and Coulomb terms \cite{blaizot80,blaizot76}. The sign of
these terms is negative and they considerably decrease the value of
$K\ur{s}_A$ in real nuclei with respect to the $K_\infty$ limit
\cite{chossy97,blaizot80,blaizot76,treiner81}. A key point is the fact
that the surface incompressibility $K\ur{s}_{\rm surf}$ in the scaling
model varies almost as a linear function of $K_\infty$, which
guarantees that the surface effects do not destroy the regular
behaviour of $K\ur{s}_A$ with $K_\infty$. For instance, our RETF
calculations of the $K\ur{s}_{\rm surf}$ coefficient for several
relativistic parameter sets \cite{patra01b}, by using the scaling
method in semi-infinite nuclear matter, confirm that $K\ur{s}_{\rm
surf}$ in the relativistic model indeed behaves roughly linearly with
the bulk compression modulus, as happens with non-relativistic Skyrme
forces. In fact we have found \cite{patra01b} that the rule of thumb
$K\ur{s}_{\rm surf} \sim - K_\infty$ known from the non-relativistic
approach \cite{treiner81}, also applies to non-linear RMF parameter
sets having not too large values of the compression modulus. In the
case of the linear $\sigma-\omega$ sets, which have a high $K_\infty$
value, one instead finds $K\ur{s}_{\rm surf} \sim -1.5 K_\infty$.

It is interesting to compare \req{eqfit0} with the equations
$K\ur{s}_A = 0.62 K_\infty + 23$ MeV for $^{208}$Pb and
$K\ur{s}_A = 0.49 K_\infty + 35$ MeV for $^{40}$Ca obtained in
Ref.\ \cite{hamamoto97} from an analysis of the scaling
incompressibility performed with the Skyrme forces SkM*, SGI and
SIII\@. In the relativistic model the independent term of the linear
fits is negative and seems to be more constant with mass number, but
the coefficient in front of $K_\infty$ is similar in both the
relativistic and non-relativistic model. The authors of Ref.\
\cite{hamamoto97} signaled that the slope obtained for $^{208}$Pb with
the Skyrme forces approaches the hydrodynamical value $\pi^2/15=
0.658$, though they stressed that this might be just accidental. It is
at least curious to come across with the same value in the
relativistic model.

In assuming a nuclear matter approach Nishizaki et al.\
\cite{nishizaki87} estimated the monopole excitation energy of a
finite nucleus in the relativistic model as
\beq
{\bar E}\dr{M}\ur{s} =
\sqrt{ \frac{K_\infty}{ \mu_\infty \langle r^2 \rangle} } ,
\label{eqFN55}\eeq
where $\mu_\infty$ is the chemical potential of nuclear matter
(including the nucleon rest mass), $\langle r^2 \rangle= \frac{3}{5}
R^2$ and $R= 1.2 A^{1/3}$ fm. They evaluated \req{eqFN55} for the
linear model of Walecka using $\mu_\infty= 923$ MeV and $K_\infty=
525$ MeV and found ${\bar E}\dr{M}\ur{s}=160/A^{1/3}$ MeV\@. This
result has the correct dependence on the mass number but it is twice
as large as the empirical value $\sim 80/A^{1/3}$ MeV\@. Calculating
\req{eqFN55} for the non-linear parametrizations NL1, NL3, NL-SH and
NL2 one finds ${\bar E}\dr{M}\ur{s}= 102$, 115, 132, and $140/A^{1/3}$
MeV, respectively. Compared to the empirical law these values are too
large by a factor $\sim 1.5$--$1.8$, depending on the compression
modulus of the force. If we furthermore compare with the results of
the calculations for finite nuclei listed in Table 1, we realize that
the finite size effects reduce the prediction obtained from nuclear
matter by a factor ranging from 1.6 in $^{16}$O to 1.3 in $^{208}$Pb,
almost independently of the parameter set.

In a recent work Piekarewicz \cite{piekarewicz01} has given a thorough
presentation of the relativistic RPA formalism and has computed the
isoscalar monopole mode for several closed-shell nuclei. In the
numerical calculations he has used the non-linear sets NLC and NLB and
the linear set L2$'$, which have the nuclear matter
incompressibilities $K_\infty= 224$, 421 and 547 MeV, respectively. We
present in Table 2 our values obtained from the scaling method versus
the RRPA peak energies of the isoscalar mode taken from Ref.\
\cite{piekarewicz01}, for the systems $^{40}$Ca, $^{90}$Zr and
$^{208}$Pb. A fairly good agreement is found between our semiclassical
calculations and the more fundamental RRPA approach. The differences
are well below 5\% in $^{208}$Pb and, excluding the RETF result for
$^{40}$Ca with L2$'$, below 10\% in $^{40}$Ca and $^{90}$Zr. As
discussed in Ref.\ \cite{piekarewicz01}, it becomes difficult to even
identify a genuine resonance in the RRPA distribution of the isoscalar
monopole strength for medium-size nuclei such as $^{40}$Ca with the
parameter sets NLB and L2$'$ which have large compression moduli.

The GMR has also been studied by means of constrained calculations in
the RMF model and, based upon them, with the more ellaborate generator
coordinate method (GCM). The constrained calculations in our
semiclassical approach (see Section \ref{constrained}) are carried out
in a similar way to that of Refs.\
\cite{maruyama89,boersma91,stoi94,stoitsov94} within the quantal
Hartree approach. We report in Table 3 the excitation energies of
$^{16}$O, $^{40}$Ca, $^{90}$Zr and $^{208}$Pb calculated with the
constrained RETF method (for the NL1, NL2, NL3 and NL-SH sets),
besides the constrained RMF (CRMF) Hartree results of Ref.\
\cite{stoi94} and the GCM results of Ref.\ \cite{vretenar97}. In
non-relativistic RPA calculations it is common to utilize the moments
$m_k$ of the strength function to analyze the monopole resonance
\cite{bohigas79}. The lowest moments correspond to simple sum rules
and in the limit of small amplitude oscillations the ratios
$\sqrt{m_3/m_1}$ and $\sqrt{m_1/m_{-1}}$ can be identified,
respectively, with the scaling and constrained monopole excitation
energies \cite{blaizot80,blaizot95,kohno79}.

We see that the excitation energies of the monopole state are smaller
in the constrained model (Table 3) than in the scaling model (Table
1). This is in agreement with the non-relativistic RPA inequality
$\sqrt{m_1/m_{-1}} \leq \sqrt{m_3/m_1}$ \cite{bohigas79}. When the
comparison is possible, the energies obtained with the GCM are
systematically smaller than in the CRMF Hartree model, which in turn
are smaller than in the constrained RETF approach. The constrained
RETF values agree very well with the GCM and CRMF values for NL1, but
the agreement worsens for light nuclei with the other parameter sets.
In the case of $^{208}$Pb, for which the semiclassical technique
should work better, the constrained RETF calculation overestimates the
GCM value by around 1 MeV, the same magnitude by which the CRMF and
GCM results differ (for NL1 and NL-SH). 

Vretenar et al.\ \cite{vretenar97} have studied the GMR with the
time-dependent RMF approach. (We recall that very recently it has been
demonstrated that the relativistic RPA, with inclusion of Dirac sea
states, is equivalent to the small amplitude limit of the
time-dependent RMF theory in the no-sea approximation \cite{ma01}.) In
the calculations of Ref.\ \cite{vretenar97} the main peak appearing in
the monopole strength distribution of $^{208}$Pb has energies 12.4
(NL1), 14.1 (NL3), 16.1 (NL-SH) and 17.8 MeV (NL2), while in the case
of $^{90}$Zr the energies are 15.7 (NL1) and $\sim 18$ MeV (NL3). For
$^{208}$Pb our scaling results (cf.\ Table 1) show a good agreement in
all the parameter sets. In fact, if we focus on the RETF values, we
see that the scaling energies are an upper bound of the time-dependent
RMF energies, while the constrained energies of Table 3 represent a
lower bound (apart from the case of NL-SH by a little deviation). For
$^{90}$Zr, however, both the scaling and constrained semiclassical
excitation energies are larger than those of Ref.\ \cite{vretenar97}.
It should be pointed out that the Fourier spectrum of $^{90}$Zr in the
time-dependent RMF calculation is considerably fragmented (specially
for the sets with higher $K_\infty$) and then the determination of the
centroid energy remains more uncertain \cite{vretenar97}.

\pagebreak

\section{Giant quadrupole resonance}
\label{quadrupole}

In the quadrupole vibration the particle density scales as
\cite{bohigas79} 
\beq
\rho_\lambda(\vect{r}) =
\rho ( x/\lambda, y/\lambda, \lambda^2 z ) .
\label{eqFN37}\eeq
While the volume element is conserved in both coordinate and momentum
space, the momentum distribution, which remained spherically symmetric
in the monopole oscillation, becomes highly deformed in the quadrupole
case \cite{ring80,jennings80}:
\beq
\vect{p}_\lambda =  (\lambda p_x, \lambda p_y, p_z/\lambda^2 ) .
\label{eqFN38}\eeq

One has to note that the spherically averaged form of the distribution
function ${\cal R}(\vect{r},\vect{p})$ cannot be employed in the
quadrupole scaling calculations due to the deformation of the Fermi
sphere \cite{jennings80}. This means, in particular, that the
spherically symmetric expressions \req{eqFN2b}, \req{eqFN2c} and
\req{eqFN8} of the semiclassical energy density and scalar density are
no longer valid for use in the quadrupole scaling. That is, first, one
should replace $\vect{p}$ by $\vect{p}_\lambda$ in the semiclassical
expansion of the relativistic distribution function ${\cal
R}(\vect{r},\vect{p})$ \cite{centelles93a} and then obtain its moments
(energy and densities) as a function of the collective coordinate
$\lambda$. This extraordinarily complicates the expressions if the
distribution function with terms up to order $\hbar^2$ is to be used.
Since the final magnitude of the contribution of the $\hbar^2$-order
corrections in the semiclassical calculation of the excitation energy
of giant resonances is not very significant, we will work at the
Thomas--Fermi level in the present study of the giant quadrupole
resonance.

In the Thomas--Fermi approach the relativistic distribution function
is proportional to a step function (Appendix A and Ref.\
\cite{centelles93a}), which vanishes for single-particle energies
above the Fermi level. The Thomas--Fermi energy density of the
non-linear $\sigma-\omega$ model after scaling then reads
\beqa
{\cal H}_\lambda & = &
\frac{2}{(2\pi)^3} \sum_{q} \int d\vect{p} \,
\sqrt{p^2_\lambda+{m^*_\lambda}^2} \,\, \Theta \left( \mu_{q \lambda}
- \sqrt{p^2_\lambda+{m^*_\lambda}^2} - u_{q \lambda} \right)
\nonumber\\[3mm]
& & \mbox{}
+\frac{1}{2}g\dr{s} \phi_\lambda \rho\ur{eff}\dr{s \lambda} 
+\frac{1}{3}b\phi_\lambda^3+\frac{1}{4}c\phi_\lambda^4 
+\frac{1}{2}g\dr{v} V_\lambda \rho_\lambda  
+\frac{1}{2}g_\rho R_\lambda \rho_{3 \lambda}
+\frac{1}{2}e {\cal A}_\lambda \rho\dr{p \lambda} ,
\label{eqFN42}
\eeqa
where $\Theta$ denotes the step function, $\mu_{q \lambda}$ is the
chemical potential of the scaled system for each kind of nucleon and
the single-particle potential $u_{q \lambda}$ is given by
\beq
u_{q \lambda } = g\dr{v} V_\lambda 
+\frac{1}{2} g_\rho R_\lambda \tau_3 
+\frac{1}{2} e {\cal A}_\lambda (1+\tau_3) .
\label{eqFN41}\eeq
The scaled effective scalar density $\rho\ur{eff}\dr{s \lambda}$
has been defined through
\beqa
g\dr{s}\rho\ur{eff}\dr{s \lambda} & = &
g\dr{s}{\rho\dr{s\lambda}}-b\phi_\lambda^2-c\phi_\lambda^3
\nonumber\\[3mm]
& = &
\frac{2}{(2\pi)^3}\sum_{q} \int d\vect{p} \,
\frac{g\dr{s} m_\lambda^*}{\sqrt{p^2_\lambda+{m^*_\lambda}^2}}
\,\, \Theta \left( \mu_{q \lambda}
- \sqrt{p^2_\lambda+{m^*_\lambda}^2} - u_{q \lambda} \right)
-b\phi_\lambda^2-c\phi_\lambda^3 .
\label{eqFN43}
\eeqa
The position and momentum variables in these expressions scale
according to the rules \req{eqFN37} and \req{eqFN38} in the quadrupole
case. 

To obtain the restoring force $C\dr{Q}$ of the quadrupole oscillation
we have to compute the second derivative of the scaled energy with
respect to the collective coordinate $\lambda$. The first derivative
reads 
\beqa & & 
\frac{\partial}{\partial \lambda}
\int d\vect{r} {\cal H}_\lambda (\vect{r})  = 
\nonumber \\[3mm] 
& &
\int d\vect{r} \left[ \frac{2}{(2\pi)^3} \sum_q \int d\vect{p} \,
\frac{p_\lambda}{\sqrt{p^2_\lambda+{m^*_\lambda}^2}}
\frac{\partial p_\lambda}{\partial \lambda}
\,\, \Theta \left( \mu_{q \lambda}
- \sqrt{p^2_\lambda+{m^*_\lambda}^2} - u_{q \lambda} \right)
- g\dr{s} \rho\ur{eff}\dr{s \lambda} 
\frac{\partial \phi_\lambda}{\partial \lambda} \right]
\nonumber \\[3mm]  
& & \mbox{} 
+ \frac{\partial}{\partial \lambda} \int d\vect{r} \left[
\frac{1}{2}g\dr{s} \phi_\lambda \rho\ur{eff}\dr{s \lambda} 
+\frac{1}{2}g\dr{v} V_\lambda \rho_\lambda
+\frac{1}{2}g_\rho R_\lambda \rho_{3 \lambda}
+\frac{1}{2}e {\cal A}_\lambda \rho\dr{p \lambda} \right] .
\label{eqFN47}\eeqa
It can be checked that this equation identically vanishes at
$\lambda=1$, as in the non-relativistic case \cite{bohigas79}. Before
deriving again \req{eqFN47} it is helpful to take into account that,
for instance,
\beqa & &
\int d \vect{r} \rho_\lambda(\vect{r}) V_\lambda(\vect{r}) =
\int d \vect{r} \rho \bigg( \frac{x}{\lambda}, \frac{y}{\lambda},
\lambda^2 z \bigg)
\int d \vect{r}' \rho \bigg( \frac{x'}{\lambda}, \frac{y'}{\lambda},
\lambda^2 z' \bigg)
\frac{g\dr{v}}{4\pi}
\frac{e^{-m\dr{v} |\vect{r}-\vect{r}'|}}{|\vect{r}-\vect{r}'|} = 
\nonumber\\[3mm]
& &
\int d \vect{r} \rho(\vect{r}) \int d \vect{r}' \rho(\vect{r}')
{\cal V}_\omega (s_\lambda) ,
\label{eqFN44}\eeqa
where
\beq
{\cal V}_\omega (s_\lambda)  = \frac{g\dr{v}}{4\pi}
\frac{e^{-m\dr{v} s_\lambda}}{s_\lambda},
\qquad
\vect{s}_\lambda = ( \lambda x - \lambda x', 
\lambda y - \lambda y', z/\lambda^2 - z'/\lambda^2 ) .
\label{eqFN45}\eeq
With this, after some algebra, the restoring force of the quadrupole
mode can be put in the form
\beqa & &
C\dr{Q} = 
\left[ \frac{\partial^2}{\partial \lambda^2} \int
 d\vect{r} {\cal H}_\lambda (\vect{r}) \right]_{\lambda=1} = 
\nonumber \\[3mm]
& &
\int d\vect{r} \Bigg\{ \frac{2}{(2\pi)^3}\sum_q \int d\vect{p}
\left[ \frac{p_x^2+p_y^2+10p_z^2}{({p^2}+{m^*}^2)^{1/2}}
-\frac{(p_x^2+p_y^2-2p_z^2)^2} {(p^2+{m^*}^2)^{3/2}} \right]
\Theta (p-p_{{\rm F}q}) 
\nonumber \\[3mm]
& & \mbox{}
+ g\dr{s} \left. \frac{\partial\rho\dr{s \lambda}\ur{eff}}
{\partial \lambda} \right|_{\lambda=1}
\int d\vect{r}' \rho\dr{s}\ur{eff}(\vect{r}')\frac{1}{s}
\frac{d {\cal V}_\sigma}{d s}s_-^2
- \frac{1}{2} g\dr{s} \rho\ur{eff}\dr{s} \int d\vect{r}'
\rho\dr{s}\ur{eff}(\vect{r}')
\left[\frac{1}{s}\frac{d}{d s} \left(\frac{1}{s}
\frac{d{\cal V}_\sigma}{d s}\right) s_-^4
+\frac{3}{s}\frac{d{\cal V}_\sigma}{d s} s_+^2\right]
\nonumber \\[3mm]
& & \mbox{}
+ \frac{1}{2} g\dr{v} \rho\int d\vect{r}'\rho(\vect{r}')
\left[\frac{1}{s}\frac{d}{d s} 
\left(\frac{1}{s}\frac{d {\cal V}_\omega} {d s}\right) s_-^4
+\frac{3}{s}\frac{d{\cal V}_\omega}{d s} s_+^2\right]
\nonumber \\[3mm]
& & \mbox{}
+\frac{1}{2} g_\rho \rho_3 \int d\vect{r}'
\rho_3(\vect{r}') \left[\frac{1}{s}\frac{d}{d s} 
\left(\frac{1}{s}\frac{d {\cal V}\dr{\rho}}{d s}\right) s_-^4
+\frac{3}{s}\frac{d{\cal V}\dr{\rho}}{d s} s_+^2\right] 
+\frac{1}{2} e \rho\dr{p}\int d\vect{r}' \rho\dr{p}(\vect{r}')
\frac{3e}{4\pi} \left( \frac{s_-^4}{s^5} - \frac{s_+^2}{s^3} \right)
\Bigg\} ,
\nonumber \\[3mm]
\label{eqFN49}\eeqa
where we have set $s_\mp^2= s_x^2 + s_y^2 \mp 2 s_z^2$.

After performing the angular average in the integral over $\vect{p}$
and in the integrals over $\vect{r}$ and $\vect{r}'$, we finally get
\beqa
C\dr{Q} & = &
\frac{2}{5} \int d\vect{r} \Bigg\{ \frac{2}{\pi^2}
\left[ \frac{k\dr{Fn}^5}{\epsilon\dr{Fn}}
     + \frac{k\dr{Fp}^5}{\epsilon\dr{Fp}} \right]
- g\dr{s} \rho\dr{s}\ur{eff}\int d \vect{r}'
\rho\dr{s}\ur{eff}(\vect{r}')
\left[ 4s\frac{d{\cal V}_\sigma}{d s}
+ s^2 \frac{d^2{\cal V}_\sigma}{d s^2} \right]
\nonumber \\[3mm]
& & \mbox{}
+ g\dr{v}\rho\int d \vect{r}' {\rho}(\vect{r}')
\left[ 4s\frac{d{\cal V}_\omega}{d s}
+ s^2 \frac{d^2{\cal V}_\omega}{d s^2} \right]
\nonumber \\[3mm]
& & \mbox{}
+ g_\rho \rho_3 \int d \vect{r}'
\rho_3(\vect{r}') \left[ 4s\frac{d{\cal V}_\rho}{d s}
+ s^2 \frac{d^2{\cal V}_\rho}{d s^2} \right] 
- 2 e {\cal A} \rho\dr{p} \Bigg\} .
\label{eqFN50}\eeqa
As far as the nuclear part is concerned this result coincides with the
one derived in Ref.\ \cite{nishizaki87} for nuclear matter using a
local Lorentz boost and the scaling method. The contributions from the
meson fields agree with the result obtained from the potential part of
an effective density-independent nuclear force in the non-relativistic
model \cite{kohno79} (and the contribution from the Coulomb field
agrees with that given in Ref.\ \cite{bohigas79}).

As in the monopole oscillation to calculate the mass parameter one
needs the continuity equation in a moving frame, Eq.\ \req{eqFN35}.
For the quadrupole vibration we have $\rho_\lambda(\vect{r}) =
\rho(x/\lambda, y/\lambda, \lambda^2 z)$ and the continuity equation
\req{eqFN35} is fulfilled by $\vect{v}= -\dot\lambda (-x,-y,2z) =
-\dot\lambda \vect{\nabla} [\sqrt{4\pi/5} r^2 Y_{20}(\Omega)]$ at
$\lambda=1$ \cite{nishizaki87}, which provides the connection between
the velocity of the moving frame and the collective coordinate.
Proceeding similarly to the monopole case, i.e., inserting this
velocity field into Eq.\ \req{eqFNA16} and taking the second
derivative with respect to $\dot\lambda$, the mass parameter of the
quadrupole mode is found to be
\beq
B\dr{Q} = 2 \int d \vect{r} r^2 {\cal H} ,
\label{eqFN53}
\eeq
assuming the nucleus to be spherical. The excitation energy of the
quadrupole state then is
\beq
{\bar E}\dr{Q}\ur{s} = \sqrt{ \frac{C\dr{Q}}{B\dr{Q}} } .
\label{eqFN53b}
\eeq
The transition density in the quadrupole case is given by
\beq
\rho_{\rm T}(\vect{r}) = \left. 
\frac{d \rho_\lambda(\vect{r})}{d \lambda} \right|_{\lambda=1}
= \left. \vect{\nabla} \rho_\lambda(\vect{r}) \right|_{\lambda=1}
\cdot \vect{\nabla} [\sqrt{4\pi/5} r^2 Y_{20}(\Omega)]
= \sqrt{\frac{16\pi}{5}} r \frac{d \rho(r)}{d r} Y_{20}(\Omega) ,
\label{eqtransq}\eeq
where again we have assumed the density to be spherically symmetric at
$\lambda= 1$.

\subsection{Numerical results}

As we have indicated, our calculations for the quadrupole mode are
restricted to the RTF approximation. We collect in Table 4 the
calculated excitation energy of the quadrupole oscillation for
$^{16}$O, $^{40}$Ca, $^{48}$Ca, $^{90}$Zr and $^{208}$Pb, along with
the empirical law $E\dr{x}\sim 65/A^{1/3}$ MeV and some experimental
data taken from Ref.\ \cite{bertrand76}. The theoretical results shown
in this table correspond to the non-linear sets NL1, NL3, NL-SH
and NL2, and to the set LZ ($K_\infty= 586$ MeV, $m^*_\infty/m=0.53$)
which we take as a representative of the linear sets.

One can see that the four non-linear $\sigma-\omega$ parametrizations
reproduce the empirical trend and that, contrary to the situation
found in the monopole case, they give rather similar results for each
nucleus. This is due to the fact that the energy of the quadrupole
vibration is basically independent of the bulk compression modulus of
the effective force. Nevertheless, the comparison with experiment
favours the NL3 set and, especially, the NL1 set (i.e., those sets
with a lower incompressibility). In fact, if the incompressiblity of
the force is very large (set LZ) the theoretical predictions clearly
overestimate the experimental values. The relativistic results of the
non-linear sets compare well with those obtained in non-relativistic
Hartree--Fock and extended Thomas--Fermi calculations using Skyrme
forces \cite{centelles90}. Calculations of the isoscalar giant
quadrupole resonance are rather scarce in the relativistic domain.
Time-dependent RMF calculations of this mode have been carried out in
Ref.\ \cite{vretenar95} using the NL-SH parameter set. Our
relativistic Thomas--Fermi calculation is in good agreement with the
excitation energies of 23.6, 17.7 and 17.7 MeV for $^{16}$O, $^{40}$Ca
and $^{48}$Ca, respectively, reported in that work.

The energy of the quadrupole excitation has also been evaluated by
Nishizaki et al.\ \cite{nishizaki87} from a nuclear matter approach as
\beq
{\bar E}\dr{Q}\ur{s} = \sqrt{ \frac{6 k^2\dr{F,\infty}}
{5 \epsilon\dr{F,\infty} \mu_\infty \langle r^2 \rangle} } ,
\label{eqQP1}\eeq
where $\langle r^2 \rangle$ has been defined in Eq.\ \req{eqFN55}. In
this approximation the restoring force of the quadrupole vibration
corresponds to the nuclear matter limit of Eq.\ \req{eqFN50}, where
all the terms with derivatives of the meson fields vanish and only the
first term survives. Note that the incompressibility $K_\infty$ of the
interaction does not enter Eq.\ \req{eqQP1}. According to this
equation one obtains ${\bar E}\dr{Q}\ur{s}= 85$, 84, 81, 80, and
$76/A^{1/3}$ MeV for the LZ, NL1, NL3, NL-SH and NL2 sets,
respectively. We thus see that in nuclear matter ${\bar
E}\dr{Q}\ur{s}$ decreases as the value of the effective mass at
saturation of the force ($m^*_\infty$) increases. However, in the full
RTF calculation for finite nuclei (Table 4) the regular pattern of
${\bar E}\dr{Q}\ur{s}$ with $m^*_\infty$ observed in nuclear matter is
destroyed by the finite size effects. In the case of finite systems
one not only has the additional contribution from the meson fields
into Eq.\ \req{eqFN50}, but also the nuclear part is modified by the
shape of the nuclear surface, this one depending in turn on the mass
of the sigma meson $m\dr{s}$. Such effects mask the simple relation of
${\bar E}\dr{Q}\ur{s}$ with $m^*_\infty$ shown by the naive nuclear
matter approximation.

\pagebreak

\section{Summary and conclusions}

We have studied the isoscalar giant monopole and quadrupole resonances
of finite nuclei by means of the scaling method and the Thomas--Fermi
and extended Thomas--Fermi approaches to relativistic mean field
theory. Self-consistent numerical calculations for realistic
non-linear $\sigma-\omega$ parameter sets have been discussed.
Previous relativistic investigations with the scaling method either
relied on a leptodermous expansion of the finite nucleus
incompressibility \cite{stoitsov94,eiff94,chossy97}, or were limited
to the linear $\sigma-\omega$ model for symmetric and uncharged nuclei
at the Thomas--Fermi level \cite{nishizaki87,zhu91}.

In the present approach one starts by scaling the spatial and momentum
coordinates of the semiclassical distribution function in a moving
frame. By taking the derivatives of the scaled energy in the moving
frame with respect to the collective coordinate and the collective
velocity, one obtains the expressions from which the restoring force
and the mass parameter of the resonance can be computed. The
underlying reason for the success of the method is that in the
semiclassical approach the energy functional is written explicitly in
terms of the local Fermi momentum {\em and} of the local effective
mass, which allows one to easily perform the scaling. Due to the
finite range of the relativistic interaction no compact formulas can
be obtained as in the case of non-relativistic Skyrme forces.
Nevertheless, the scaling excitation energies of the monopole and
quadrupole resonances only depend on the ground-state densities and
fields, which means that they can be computed as a by-product of a
semiclassical self-consistent calculation of the ground state.

We have found that the total contribution to the excitation energy of
the GMR coming from the gradient corrections of order $\hbar^2$, which
are included in the RETF approach, does not modify the Thomas--Fermi
result very much. The strength and sign of these corrections of order
$\hbar^2$ is strongly correlated with the nuclear matter
incompressibility and the effective mass at saturation of the
relativistic interaction.

We have investigated the relation between the incompressibility
$K\ur{s}_A$ of finite nuclei in the scaling model and the
compression modulus of nuclear matter $K_\infty$, employing a variety
of relativistic parameter sets. The dependence is roughly linear, as
in non-relativistic analyses. Even a nucleus such as $^{208}$Pb is not
large enough to obtain a relation of proportionality between
$K\ur{s}_A$ and $K_\infty$. The excitation energy of the monopole
oscillation increases smoothly with $K_\infty^{1/2}$, in
correspondence with the behaviour of $K\ur{s}_A$. No regular
pattern of the monopole excitation energy with the mass of the scalar
meson or with the effective mass of the interaction has been observed.

The experimental excitation energies of the monopole oscillation in
medium and heavy nuclei lie in between the results obtained with the
NL1 and NL3 parameter sets. An analysis of the calculated
breathing-mode energies for $^{116}$Sn, $^{144}$Sm and $^{208}$Pb, for
which precise experimental data exist, predicts that the nuclear
matter incompressibility should be around $220$--$260$ MeV
($230$--$260$ MeV if only $^{144}$Sm and $^{208}$Pb are taken into
account). A similar analysis carried out in Refs.\
\cite{vretenar97,ma01} using time-dependent RMF and relativistic RPA
results predicts a value slightly higher: 250--270 MeV\@. From the
relativistic RPA peak energies given in Ref.\ \cite{piekarewicz01} for
$^{208}$Pb we extract a range of 235--250 MeV\@. Thus, all these
relativistic calculations point to a value of roughly $250\pm20$ MeV
for $K_\infty$, which is higher than the non-relativistic estimate of
$215\pm15$ MeV from Skyrme and Gogny forces \cite{blaizot95,farine97}.
Relativistic parameter sets with large values of $K_\infty$ (such as
NL-SH or NL2), which may otherwise perform well in describing the data
for nuclear masses and radii, should be discarded on the basis of the
experimental information on breathing-mode energies.

The results computed with the scaling method represent an upper bound
of the mean excitation energy of the GMR, to the extent that they are
related with the cubic weighted sum rule. Instead, the breathing-mode
energies obtained from constrained calculations rather represent a
lower bound, since they are related with the inverse energy-weighted
sum rule. Actually, with all the parameter sets and nuclei analyzed,
we have found the calculated monopole energies to be larger in the
scaling approach than in the constrained approach.

Our calculations of the excitation energy of the quadrupole
oscillation have been restricted to the Thomas--Fermi approach, to
simplify the problems related with the distortion of the Fermi sphere.
All the considered non-linear parameter sets reproduce fairly well the
empirical trend, rather independently of the value of the compression
modulus of the force. Although a nuclear matter estimate predicts a
decrease of the quadrupole excitation energy with an increase in the
value of the effective mass at saturation, the finite size effects and
additional contributions from the meson fields mask this trend in the
self-consistent calculations for actual nuclei.

In conclusion, we hope to have shown that the scaling method can be
confidently used together with the relativistic Thomas--Fermi approach
to estimate the excitation energy of the isoscalar monopole and
quadrupole resonances in a simple and reliable way. The results for
the breathing mode turn out to be in good agreement with the outcome
of dynamical time-dependent RMF and relativistic RPA calculations. We
can thus conclude that, similarly to the non-relativistic case, also
in the relativistic framework the semiclassical excitation energies
obtained with the scaling method simulate the results of the RPA\@.
The method introduced in this work also allows one to
self-consistently compute the surface incompressibility coefficient
for relativistic interactions \cite{patra01b}. The study of other
multipolarities using a generalized scaling simultaneously with the
relativistic Thomas--Fermi approach may be a worthwhile task to
pursue.

\section{Acknowledgements}

We thank Nguyen Van Giai for discussions and J. Navarro for carefully
reading an earlier version of the manuscript and making a series of
suggestions. Support from the DGICYT (Spain) under grant PB98-1247 and
from DGR (Catalonia) under grant 2000SGR-00024 is acknowledged. S.K.P.
thanks the Spanish Education Ministry grant SB97-OL174874 for
financial support and the Departament d'Estructura i Constituents de
la Mat\`eria of the University of Barcelona for kind hospitality.

\pagebreak

\renewcommand {\theequation}{\Alph{section}.\arabic{equation}}
\setcounter{section}{1}
\setcounter{equation}{0}
\section*{Appendix A}

In this appendix we derive the Thomas--Fermi expression of the energy
of a nucleus described by the non-linear $\sigma-\omega$ model in a
frame moving with velocity $-\vect{v}$. As a final product we obtain
general equations for the restoring force and the mass parameter of
the giant resonance. For simplicity we shall consider an uncharged
symmetric nucleus (the $\rho$ meson field and the electromagnetic
field behave like the vector field), and shall not include the
corrections of order $\hbar^2$ to the Thomas--Fermi approximation.

The semiclassical expressions of densities and energies are most
conveniently derived from the so-called phase-space distribution
function \cite{ring80}. For a Hamiltonian $\vect{\alpha} \cdot
\vect{p} + \beta m^* + g\dr{v} V$ the distribution function in
Thomas--Fermi approximation reads \cite{centelles93a}
\beq
{\cal R} = \frac{1}{2} \Theta (\mu-\epsilon- g\dr{v} V) 
\left[ I + \frac{1}{\epsilon} \{ \vect{\alpha} \cdot \vect{p}
+ \beta m^* \} \right] ,
\label{eqFNA5}\eeq
where $\mu$ is the chemical potential, $\epsilon=\sqrt{p^2+{m^*}^2}$
and $I$ is the $4\times4$ unit matrix. Due to the step function in
\req{eqFNA5}, $p$ takes values from zero up to the Fermi momentum
$p\dr{F}$. The scalar field ($\phi$) and the time-like component of
the vector field ($V$) transform to a frame which moves with velocity
$-\vect{v}$ like $\phi' = \phi$ and $V' = \gamma V$, with $\gamma=
1/\sqrt{1-v^2}$. The distribution function in the moving frame then is
given by
\beq
{\cal R}' = \frac{1}{2} \Theta (\mu'-\epsilon'-\gamma g\dr{v} V)
\left[ I+ \frac{1}{\epsilon'} \{ \vect{\alpha} \cdot
(\vect{p}'- g\dr{v}\vect{V}')+ \beta m^* \} \right] ,
\label{eqFNA6}\eeq
where $\mu'$ is the chemical potential in the new frame, $\epsilon'=
\sqrt{(\vect{p}'-g\dr{v} \vect{V}' )^2+{m^*}^2}$, and we have defined
\beq
\vect{V}' \equiv \gamma \, \vect{v} V .
\label{eqFNA0}\eeq
It is easy to see that $\Theta (\mu'-\epsilon'-\gamma g\dr{v} V ) =
\Theta (\mu-\epsilon-g\dr{v} V)$ [$\mbox{} = \Theta (p\dr{F}-p)$] by
expressing $\epsilon'$ and $\mu'$ through their values in the rest
frame: 
\beq
 \epsilon'  =  \gamma ( \epsilon + \vect{p} \cdot \vect{v} ) ,
\qquad
 \mu'  =  \gamma ( \mu + \vect{p} \cdot \vect{v} ) .
\label{eqFNA8}\eeq

The baryon density in the moving frame is obtained as
\beq
\rho' = 2 \int \frac{d\vect{p}'}{(2\pi)^3} 
{\rm T\/r} [\, {\cal R}' \,]  = 4 \int \frac{d\vect{p}}{(2\pi)^3}
\frac{\gamma}{\epsilon} ( \epsilon + \vect{p} \cdot \vect{v} )
\Theta(p\dr{F}-p) = \gamma \rho ,
\label{eqFNA9}\eeq
where we have taken into account that $d\vect{p}'/\epsilon' =
d\vect{p}/\epsilon$ (Lorentz scalars) and the fact that the trace of
the distribution function ${\cal R}'$ equals $2 \Theta(p\dr{F}-p)$.
Similarly, the transformed scalar density is
\beq
\rho\dr{s}' = 2 \int \frac{d\vect{p}'}{(2\pi)^3}
{\rm T\/r} [\, \beta {\cal R}' \,]
= 4 \int \frac{d\vect{p}}{{(2\pi)}^3}
\frac{m^*}{\epsilon} \Theta(p\dr{F}-p) = \rho\dr{s} .
\label{eqFNA10}\eeq
The energy density in the moving frame is given by 
\beqa
{\cal H}' & = & 2 \int \frac{d\vect{p}'}{(2\pi)^3}
{\rm T\/r} [\, H' {\cal R}' \,]
+ \frac{1}{2} \left[ (\vect{\nabla}' \phi)^2+m\dr{s}^2\phi^2\right]
+ \frac{1}{3}b \phi^3 + \frac{1}{4}c \phi^4
\nonumber \\[3mm]
& & \mbox{}
- \frac{1}{2} \left[ \gamma^2 (\vect{\nabla}' V)^2
+ \gamma^2 m\dr{v}^2 V^2 - (\vect{\nabla}' \times \vect{V}')^2
- m\dr{v}^2 {\vect{V}'}^2 \right] ,
\label{eqFNA11}\eeqa
where $H' = \vect{\alpha} \cdot ( \vect{p}'- g\dr{v} \vect{V}') +
\beta m^*+\gamma g\dr{v} V$.

If spherical symmetry of the meson fields is assumed Eq.\
\req{eqFNA11} becomes
\beqa
{\cal H}' & = & 4\int \frac{d\vect{p}}{(2\pi)^3}
\frac{\gamma^2}{\epsilon} ( \epsilon + \vect{p} \cdot \vect{v} )
 (\epsilon+ \vect{p} \cdot \vect{v} + g\dr{v} V ) \Theta(p\dr{F}-p)
\nonumber \\[3mm]
& & \mbox{}
+\frac{1}{2}\left[(\vect{\nabla}\phi)^2 + m\dr{s}^2\phi^2
+\frac{2}{3} \gamma^2 v^2 (\vect{\nabla}\phi)^2 \right]
+\frac{1}{3}b\phi^3+\frac{1}{4}c\phi^4
\nonumber \\[3mm]
& & \mbox{}
-\frac{1}{2}\left[ (\vect{\nabla} V)^2 + m\dr{v}^2 V^2
+\frac{2}{3} \gamma^2 v^2 (\vect{\nabla} V)^2 \right] .
\label{eqFNA12}\eeqa
After integration over momentum, the relativistic energy density in
the moving frame can be written as
\beq
{\cal H}' = \gamma^2 \left\{ {\cal H} + v^2 \left[ \rho\epsilon\dr{F}
+g\dr{v}\rho V +\frac{1}{3} (\vect{\nabla}\phi)^2
-\frac{1}{3} (\vect{\nabla} V)^2 - {\cal H} \right] \right\} ,
\label{eqFNA13}\eeq
where ${\cal H}$ is the energy density in the rest frame (Section
\ref{energy}). Equation \req{eqFNA13} agrees with the transformation
law of the stress tensor as discussed in Ref.\ \cite{baym76}. For a
uniform system ($\vect{\nabla}\phi=\vect{\nabla} V=0$) it also
coincides with the result obtained in Ref.\ \cite{nishizaki87} from a
local Lorentz boost.

Finally, the energy of the system in the moving frame is obtained by
integrating \req{eqFNA13} over the space. Taking into account the
Lorentz contraction of the volume element, this yields
\beq
E(v) = \int \frac{d\vect{r}}{\gamma} {\cal H}'
= \int d\vect{r} \gamma
\left\{ (1-v^2) {\cal H}+v^2
\left[\rho\epsilon\dr{F}+g\dr{v}\rho
V +\frac{1}{3}\left(\vect{\nabla}\phi\right)^2
-\frac{1}{3}\left(\vect{\nabla} V\right)^2\right]\right\} .
\label{eqFNA14} \eeq
Combining this result with the meson field equations and the virial
theorem derived in Appendix B, Eq.\ \req{eqFN21d}, the energy in the
new frame reads
\beq
E(v) = \int d\vect{r} \gamma \{ (1-v^2) {\cal H} + v^2 {\cal H} \}
     = \int d\vect{r} \gamma {\cal H} .
\label{eqFN33}\eeq

The restoring force of the monopole and quadrupole oscillations is
obtained by appropriately scaling the densities and mean fields in
Eq.\ \req{eqFN33} and then computing the second derivative at $v=0$
and $\lambda=1$:
\beq
C = \left[ \frac{\partial^2}{\partial\lambda^2}
\int \frac{d\vect{r}}{\gamma} {\cal H}_\lambda^\prime
\right]_{v=0,\lambda=1}
= \left[\frac{\partial^2}{\partial\lambda^2}
\int d\vect{r} {\cal H}_\lambda \right]_{\lambda=1} ,
\label{eqFNA15}\eeq
where ${\cal H}_\lambda^\prime$ and ${\cal H}_\lambda$ denote the
scaled energy densities in the moving and rest frames, respectively.
The mass or inertia parameter of the giant resonance is furnished by
the second derivative of the scaled energy in the moving frame with
respect to $\dot\lambda=d\lambda/dt$:
\beq
 B = \left[ \frac{\partial^2}{\partial \dot\lambda^2}
\int \frac{d\vect{r}}{\gamma} {\cal H}_\lambda^\prime
\right]_{\dot\lambda=0,\lambda=1} 
= \left[ \frac{\partial^2}{\partial \dot\lambda^2} \int d\vect{r} 
\gamma {\cal H}_\lambda \right]_{\dot\lambda=0,\lambda=1} .
\label{eqFNA16}\eeq
To evaluate \req{eqFNA16} it is necessary to relate the velocity
$\vect{v}$ of the moving frame with the collective velocity
$\dot\lambda$. This is achieved by scaling the continuity equation
\beq
\frac{\partial}{\partial t} \int \frac{d\vect{p}'}{\gamma (2\pi)^3}
{\rm T\/r} [\, {\cal R}' \,]
+\vect{\nabla} \cdot \int \frac{d\vect{p}'}{\gamma (2\pi)^3}
{\rm T\/r} [\, \vect{\alpha} {\cal R}' \,] = 0 ,
\label{eqFN34}\eeq
which after some algebra results into
\beq
\frac{\partial \rho_\lambda}{\partial t} +  \vect{\nabla}
\cdot ( \vect{v} \rho_\lambda ) = 0 ,
\label{eqFN35}\eeq
in terms of the scaled baryon density $\rho_\lambda$. Once the scaling
law of the baryon density with the $\lambda$ parameter is specified,
Eq.\ \req{eqFN35} will provide the connection between the velocity
$\vect{v}$ and $\dot\lambda$.

\pagebreak

\setcounter{section}{2}
\setcounter{equation}{0}
\section*{Appendix B}

The virial theorem results from homogeneity properties of the kinetic
energy and potential energy components of the energy with respect to a
scaling transformation that preserves the normalization \cite{parr89}.
For example, the scaling method has been employed to derive the virial
theorem for the Skyrme interaction \cite{bohigas79}, or for
relativistic particles bound in vector and scalar potentials
\cite{brack83}. Concerning the relativistic model discussed in the
present work, we have given the expression of the first derivative of
the scaled energy with respect to the scaling parameter $\lambda$ in
Eq.\ \req{eqFN19} of Section \ref{restoring}. It must vanish at
$\lambda=1$ (virial theorem):
\beq
0 = \left[ \frac{\partial}{\partial\lambda} \int d (\lambda \vect{r}) 
\frac{ {\cal H}_\lambda (\vect{r})}{\lambda^3} \right]_{\lambda=1} .
\label{eqFN1999}\eeq

To evaluate the above equation knowledge of the derivatives of the
scaled fields with respect to $\lambda$ is required. Starting with
the omega field $V_\lambda$, it fulfils the Klein--Gordon equation
\beq
 (\Delta - m\dr{v}^2) V_\lambda(\vect{r}) =
     -g\dr{v} \rho_\lambda(\vect{r}) ,
\label{eqFN21a}
\eeq
whose solution is
\beq
V_\lambda(\vect{r}) = \frac{g\dr{v}}{4\pi}
\int d \vect{r}' \rho_\lambda(\vect{r}')
\frac{e^{-m\dr{v} |\vect{r}-\vect{r}'|}}{|\vect{r}-\vect{r}'|}
= \int d (\lambda\vect{r}') \rho(\lambda\vect{r}') 
{\cal V}_\omega (s) .
\label{eqFN21}
\eeq
We employ the notation ${\cal V}_\omega (s) = g\dr{v} \exp{(-m\dr{v}
s)} / 4\pi s$, with $s=|\vect{r}-\vect{r}'|$, as in the main text. On
defining $\vect{u}= \lambda\vect{r}$ and $\vect{u}'= \lambda\vect{r}'$
one obtains $V_\lambda(\vect{r})= \int d \vect{u}' \rho(\vect{u}')
{\cal V}_\omega (|\vect{u}-\vect{u}'|/\lambda)$, whence
\beq
\left. \frac{\partial V_\lambda(\vect{r})}{\partial \lambda}
\right|_{\lambda=1} 
= -\int d\vect{r}' \rho(\vect{r}')
\, s \frac{d {\cal V}_\omega(s)}{d s} ,
\label{eqFN23}
\eeq
in agreement with the result given in Ref.\ \cite{kohno79}. Analogous
results are found for the scaled rho and Coulomb fields. The result
for the scalar field is more complicated because an additional term
appears due to the fact that the density ${\tilde\rho}\ur{eff}\dr{s}$
itself is a function of $\lambda$:
\beq
\left. \frac{\partial \phi_\lambda(\vect{r})} 
{\partial \lambda}\right|_{\lambda=1}
= - \int d \vect{r}'\rho\dr{s}\ur{eff} (\vect{r}') s
\frac{d{\cal V}_\sigma(s)}{d s} 
+ \int d \vect{r}' {\cal V}_\sigma (s) 
\left[ \frac{\partial{\tilde\rho}\ur{eff}\dr{s} (\lambda\vect{r}')}
{\partial \lambda} \right]_{\lambda=1} .
\label{eqFN24}\eeq
Since $g\dr{s}{\tilde\rho}\ur{eff}\dr{s} = g\dr{s} {\tilde\rho}\dr{s}
- b \phi_\lambda^2 / \lambda^3 - c \phi_\lambda^3 / \lambda^3$, we
have 
\begin{equation}
g_{\rm s} \left.  \frac{\partial{\tilde\rho}^{\rm eff}_{\rm s}}
{\partial \lambda} \right|_{\lambda=1} 
=
g_{\rm s} \left.  \frac{\partial{\tilde\rho}_{\rm s}}
{\partial \lambda} \right|_{\lambda=1} 
+ 3 ( b \phi^2 + c \phi^3 ) - ( 2 b \phi + 3 c \phi^2 )
\left. \frac{\partial \phi_\lambda}
{\partial \lambda}\right|_{\lambda=1}, 
\label{eqFN244}\end{equation}
with
\begin{equation}
\left.  \frac{\partial{\tilde\rho}_{\rm s}}
{\partial \lambda} \right|_{\lambda=1}  =
\left. \frac{\delta{\tilde\rho}_{\rm s}}{\delta {\tilde m}^*}
\frac{\partial {\tilde m}^*}{\partial \lambda} \right|_{\lambda=1}  =
 - \frac{\delta \rho_{\rm s}}{\delta m^*}
\left[ m^* + g_{\rm s} \frac{\partial \phi_\lambda}
{\partial \lambda} \right]_{\lambda=1} ,
\label{eqFN245}\end{equation}
cf.\ Eq.\ \req{eqFN20} for $\partial {\tilde m}^* / \partial \lambda$.

From substitution into Eq.\ (\ref{eqFN1999}) of the derivatives
(\ref{eqFN23}) and (\ref{eqFN24}) and of the corresponding results for
the rho and Coulomb fields, on account of Eq.\ (\ref{eqFN19}), one
obtains 
\beqa
0  & = & \int d \vect{r} 
\left[ {\cal E} - m^* \rho\dr{s}
+\frac{1}{2}g\dr{s}\rho\dr{s}\ur{eff}
\int d \vect{r}'\rho\dr{s}\ur{eff}(\vect{r}')
s \frac{d{\cal V}_\sigma}{d s} 
-b\phi^{3}-\frac{3}{4}c\phi^{4} \right.
\nonumber \\[3mm]
& & \left. \mbox{}
-\frac{1}{2}g\dr{v} \rho \int d \vect{r}'
\rho(\vect{r}') s \frac{d{\cal V}_\omega}{d s}
-\frac{1}{2}g_\rho \rho_3 \int d \vect{r}' \rho_3(\vect{r}') s
\frac{d{\cal V}\dr{\rho}}{d s} +\frac{1}{2} e {\cal A} \rho\dr{p}
\right] . 
\label{eqFN25} \eeqa
Now, using for example the relation $s \, d{\cal V}_\omega/d s = -
{\cal V}_\omega - m\dr{v} s {\cal V}_\omega$, it can be verified that
\begin{equation}
-\frac{1}{2} \int d \vect{r} g\dr{v} \rho \int d \vect{r}'
\rho(\vect{r}') s \frac{d{\cal V}_\omega}{d s}
= \int d \vect{r} \left[  \frac{1}{2} g_{\rm v} \rho V +
 m_{\rm v}^2 V^2 \right] .
\label{eqv3}\end{equation}
After similar straightforward manipulations with the other fields, the
virial theorem for the non-linear $\sigma-\omega$ model finally
becomes 
\beqa
0  & = & \int d \vect{r} 
\left[ {\cal E} - m^* \rho\dr{s}
- \frac{1}{2}g\dr{s} \phi \rho\dr{s} - m\dr{s}^2 \phi^2
- \frac{1}{2}b\phi^3 - \frac{1}{4} c \phi^4 \right.
\nonumber \\[3mm]
& & \left. \mbox{}
+\frac{1}{2} g\dr{v}  V \rho + m\dr{v}^2 V^2
+\frac{1}{2} g_\rho R \rho_3 + m\dr{\rho}^2 R^2
+\frac{1}{2} e {\cal A} \rho\dr{p} \right] .
\label{eqFN21d}\eeqa
One may notice that the quantity ${\cal E} - m^* \rho\dr{s}$
corresponds to the semiclassical average of $\sum_i
\varphi_i^{\dagger} \vect{\alpha} \cdot \vect{\nabla} \varphi_i$.
Actually, in terms of the kinetic energy density $\tau$ (namely, the
semiclassical counterpart of $\sum_i \varphi_i^{\dagger} [
\vect{\alpha} \cdot \vect{\nabla} + \beta m - m ] \varphi_i$) we can
write ${\cal E} - m^* \rho_{\rm s} = \tau + m \rho- m \rho_{\rm s}$,
which makes more obvious the kinetic energy component in the virial
theorem. In the limit of symmetric infinite nuclear matter Eq.\
\req{eqFN21d} goes over
\beq
{\cal E}_{0,\infty} - m^*_\infty \rho\dr{s,\infty}
-\frac{3}{2}\frac{g\dr{s}^2}{m^2\dr{s}}
{\rho\ur{eff}\dr{s,\infty}}^2
-b\phi^3_\infty-\frac{3}{4}c\phi^4_\infty 
+\frac{3}{2}\frac{g\dr{v}^2}{m\dr{v}^2} \rho_\infty^2 =  3 P = 0 ,
\label{eqFN26}\eeq
with $P$ being the pressure, if equilibrium quantities are used.

Taking advantage of Eq.\ (\ref{eqFN21d}) to eliminate $\int d \vect{r}
\cal E$ from the expression of $\int d \vect{r} \cal H$, the energy of
a nucleus in the RMF model takes the remarkably simple form
\begin{equation}
\int d \vect{r}  [ {\cal H} - m \rho ] =
\int d \vect{r}  \left[
m (\rho_{\rm s} - \rho) + m_{\rm s}^2 \phi^2
+ \frac{1}{3} b\phi^3 - m_{\rm v}^2 V^2 - m_\rho^2 R^2 \right] ,
\label{eqFN21h}\end{equation}
where we have subtracted the nucleon rest mass contribution. This
expression displays very clearly the relativistic mechanism for
nuclear binding. It stems from the cancellation between the scalar and
vector potentials and from the difference between the scalar density
and the baryon density (i.e., from the small components of the wave
functions). We have verified that Eqs.\ (\ref{eqFN21d}) and
(\ref{eqFN21h}) are satisfied not only by the Thomas--Fermi solutions,
but also by the ground-state densities and meson fields obtained from
a quantal Hartree calculation. Of course, the energy stationarity
condition of the RMF model against dilation must be fulfilled by any
approximation scheme utilized to solve the problem.

\pagebreak


%
\pagebreak

%
%
\section*{Figure captions}
\begin{description}
\item[Figure 1.]
The breathing-mode energies from the RETF scaling calculations are
compared with the empirical law $80/A^{1/3}$ MeV and the experimental
data reported in Ref.\ \cite{shlomo93}, as a function of the size of
the nucleus. 
\item[Figure 2.]
Scaling incompressibility of some finite nuclei as obtained in the
RETF calculations versus the nuclear matter incompressibility for
various relativistic parameter sets. The value of $K_\infty$ of each
set is listed in MeV\@. The straight lines are linear fits. The fit
for $^{208}$Pb is drawn by a solid line.
\item[Figure 3.]
Monopole excitation energy from RETF scaling calculations versus the
nuclear matter incompressibility for various relativistic parameter
sets. The value of $K_\infty$ of each set is given in Figure 2. The
dashed lines are linear fits to the square root of $K_\infty$.
\item[Figure 4.]
Monopole excitation energy of $^{208}$Pb from RETF scaling
calculations, as a function of the mass of the scalar meson and of the
nuclear matter effective mass of the relativistic interaction.
The dashed lines show the sense of increasing $K_\infty$.
\end{description}

\pagebreak

%
\begin{table}
\begin{center}
\caption{Excitation energy of the monopole state (in MeV) obtained in
the scaling approach by using various relativistic parameter sets (in
order of increasing value of the compression modulus $K_\infty$).
The experimental centroid energies are from Ref.\ \cite{young99}.}
\vspace{0.5cm}
\begin{tabular}{rcccccccccc}
\hline
 & \mc{2}{c}{NL1} & \mc{2}{c}{NL3} & \mc{2}{c}{NL-SH} & \mc{2}{c}{NL2}\\
 & RTF& RETF &RTF &RETF &RTF &RETF &RTF & RETF& $80A^{-1/3}$&Exp. \\
\hline
$^{16}$O  &25.1&23.3 & 27.5 &27.8 & 30.7& 33.1& 34.4 & 35.6 & 31.7 &\\
$^{40}$Ca &21.2&20.6 & 23.5 &24.1 & 26.6& 28.2& 29.5 & 30.3 & 23.4 & $19.2\pm0.4$\\
$^{48}$Ca &20.0&19.5 & 22.3 &22.7 & 25.2& 26.5& 27.7 & 28.3 & 22.0 &\\
$^{90}$Zr &17.2&16.9 & 19.2 &19.5 & 21.9& 22.8& 24.0 & 24.5 & 17.9 & $17.9\pm0.2$\\
$^{116}$Sn&15.9&15.6 & 17.7 &18.0 & 20.3& 21.0& 22.3 & 22.6 & 16.4 & $16.1\pm0.1$\\
$^{144}$Sm&14.9&14.6 & 16.6 &16.8 & 19.0& 19.6& 20.8 & 21.1 & 15.3 & $15.4\pm0.3$\\
$^{208}$Pb&12.9&12.7 & 14.5 &14.6 & 16.6& 17.0& 18.1 & 18.4 & 13.5 & $14.2\pm0.3$\\
\hline                
\end{tabular}
\end{center}
\end{table}

%
\begin{table}
\begin{center}
\caption{Comparison of the giant monopole resonance energies (in MeV)
obtained in the scaling model with those obtained in the relativistic
RPA \cite{piekarewicz01}.}
\vspace{0.5cm}
\begin{tabular}{rccccccccc}
\hline
  & \mc{3}{c}{NLC}  & \mc{3}{c}{NLB} & \mc{3}{c}{L2$'$} \\
  & RTF& RETF& RRPA& RTF& RETF& RRPA& RTF& RETF& RRPA \\
\hline
$^{40}$Ca & 22.5& 22.4 & 21.0& 27.7 & 29.4& 27.9& 29.1 & 33.0 & 27.3 \\
$^{90}$Zr & 18.1& 18.1 & 16.9& 23.3 & 24.2& 24.1& 25.2 & 27.4 & 26.5 \\
$^{208}$Pb& 13.6& 13.5 & 13.1& 18.0 & 18.5& 18.1& 19.9 & 21.0 & 20.1 \\
\hline
\end{tabular}
\end{center}
\end{table}

\newpage

%
\begin{table}
\begin{center}
\caption{Monopole excitation energy (in MeV) obtained by constrained
calculations with various parameter sets. The constrained RMF results
are from Ref.\ \cite{stoi94} and the generator coordinate method
results are from Ref.\ \cite{vretenar97}.}
\begin{tabular}{rcccccccccc}
\hline
 & \mc{3}{c}{NL1} & \mc{2}{c}{NL3}  & \mc{3}{c}{NL-SH} & \mc{2}{c}{NL2} \\
 & RETF&CRMF&GCM&RETF&GCM & RETF&CRMF&GCM &RETF&GCM \\
\hline
$^{16}$O  &21.8 &20.9 &20.2 &26.0 &22.6 & 30.0& 25.8& 25.0&32.4 &27.1 \\
$^{40}$Ca &19.8 &19.2 &16.6 &23.2 &19.6 & 26.9& 23.9& 22.0&29.0 &24.4 \\
$^{90}$Zr &16.5 &16.3 &14.1 &19.1 &16.9 & 22.1& 21.1& 19.5&23.7 &21.9 \\
$^{208}$Pb&12.1 &12.2 &11.0 &14.0 &13.0 & 16.2& 16.1& 15.0&17.4 &16.6 \\
\hline
\end{tabular}
\end{center}
\end{table}

%
\begin{table}
\begin{center}
\caption{Excitation energy of the quadrupole vibration (in MeV)
obtained in the scaling approach. The experimental values are from
Ref.\ \cite{bertrand76}.}
\vspace{0.5cm}
\begin{tabular}{ccccccccccc}
\hline
          & NL1  & NL3 & NL-SH &NL2  & LZ   &$65A^{-1/3}$&Exp. \\
\hline
$^{16}$O   &21.6 &22.9 &24.0   &24.7 &25.8  &25.8 &22.0\\
$^{40}$Ca  &17.9 &18.6 &19.2   &19.4 &20.8  &19.0 &18.0 \\
$^{48}$Ca  &16.9 &17.5 &18.1   &18.1 &19.5  &17.9 &    \\
$^{90}$Zr  &14.4 &14.8 &15.2   &15.1 &16.4  &14.5 &14.5\\
$^{208}$Pb &10.9 &11.2 &11.5   &11.3 &12.4  &11.0 &10.5\\
\hline
\end{tabular}
\end{center}
\end{table}

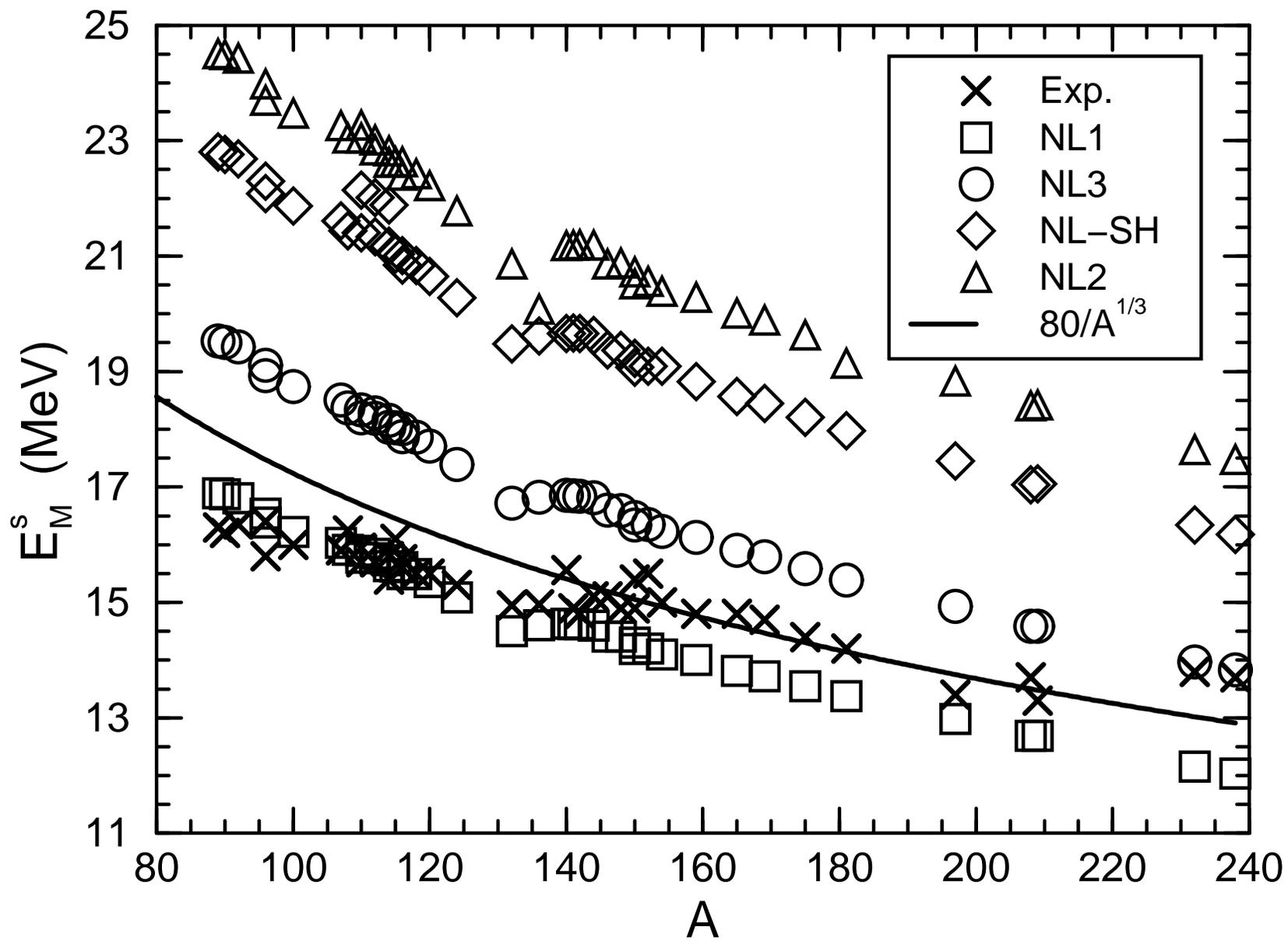
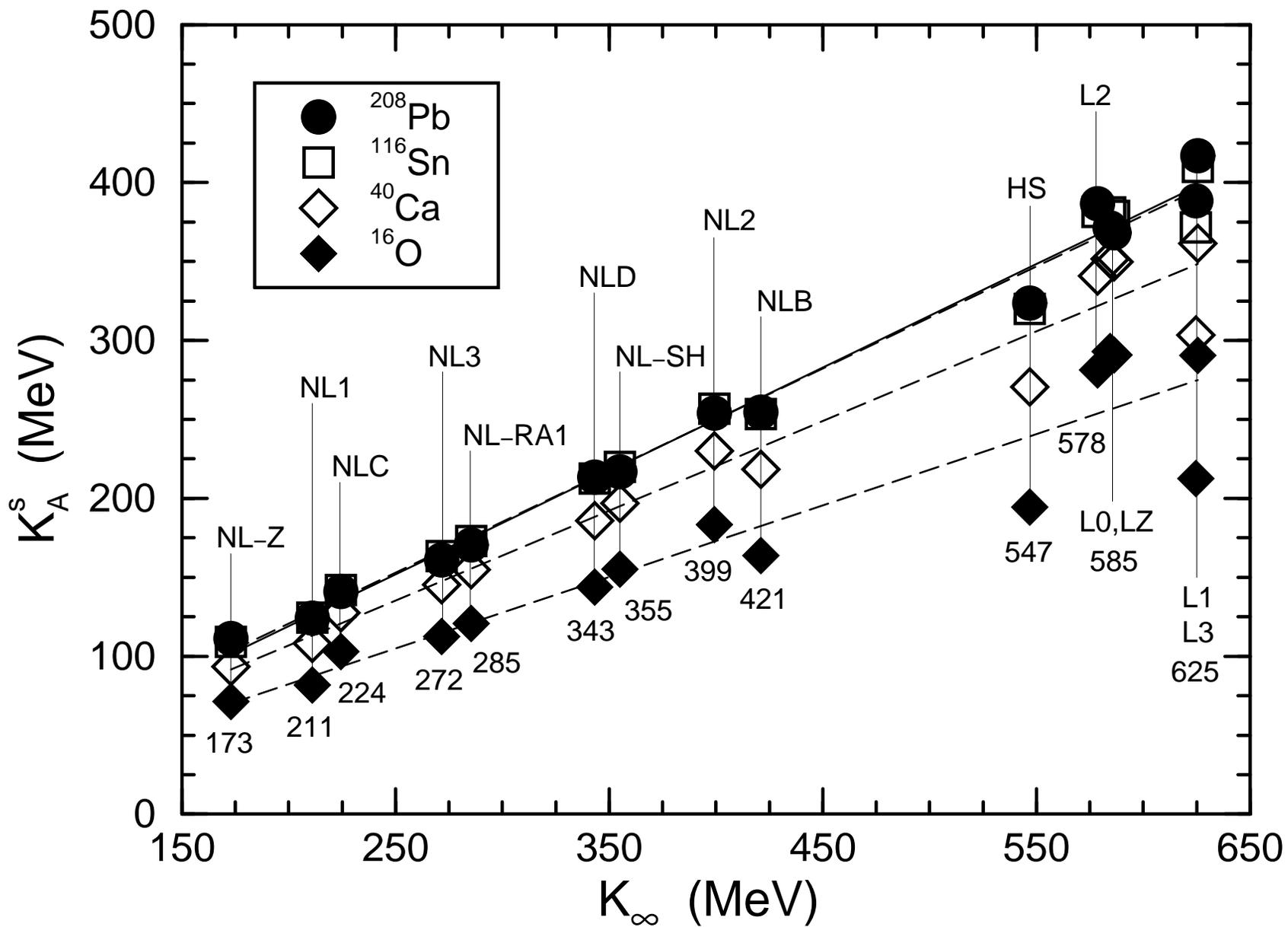
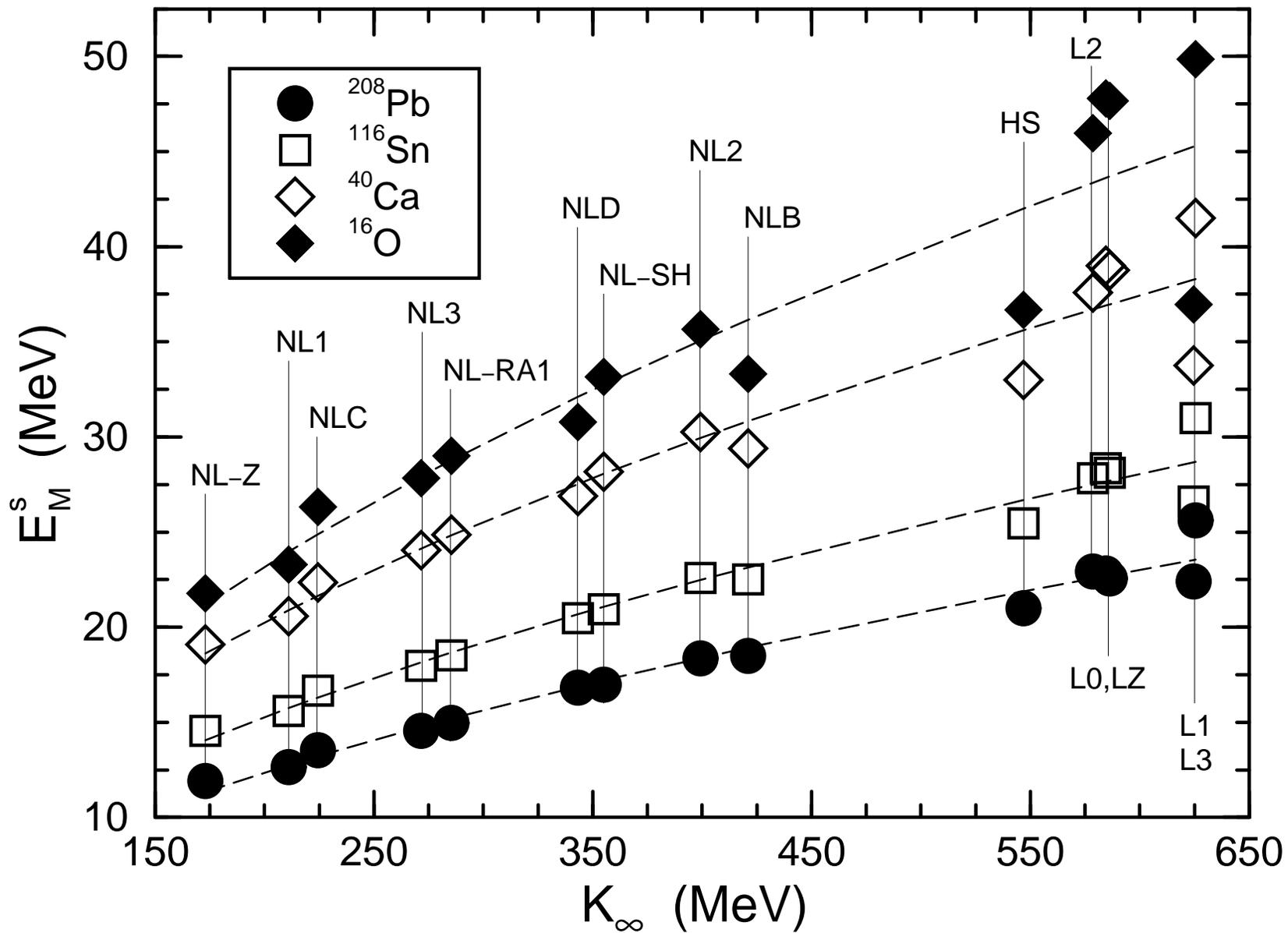
\end{document}